\documentclass[twocolumn]{aastex631}
\usepackage[T1]{fontenc}
\usepackage[normalem]{ulem}
\usepackage[utf8]{inputenc}
\shorttitle{Witnessing accretion on the tight stellar companion to HT Lup with VLT/MUSE}
\shortauthors{Jorquera et al.}

\begin{document}

\title{VLT/MUSE detection of accretion-ejection associated with \\
the close stellar companion in the HT Lup system}

\author[0000-0001-6822-7664]{Sebasti\'an Jorquera}
\affiliation{Departamento de Astronom\'ia, Universidad de Chile, Camino El Observatorio 1515, Las Condes, Santiago, Chile}
\affiliation{Universit\'e Cote d’Azur, 06304 Nice, France}
\author[0000-0001-5579-5339]{Mickaël Bonnefoy} 
\affiliation{Universit\'e Grenoble Alpes, CNRS, IPAG, 38000 Grenoble, France}
\author[0000-0002-1199-9564]{Laura M. P\'erez}
\affiliation{Departamento de Astronom\'ia, Universidad de Chile, Camino El Observatorio 1515, Las Condes, Santiago, Chile}
\author[0000-0003-4022-8598]{Gaël Chauvin}
\affiliation{Laboratoire Lagrange, Universit\'e Cote d’Azur, CNRS, Observatoire de la Cote d’Azur, 06304 Nice, Francey}
\affiliation{Unidad Mixta Internacional Franco-Chilena de Astronom\'ia, CNRS/INSU UMI 3386}
\author{Adrian Aguinaga}
\affiliation{Universit\'e Grenoble Alpes, CNRS, IPAG, 38000 Grenoble, France}
\author[0000-0001-6660-936X]{Catherine Dougados}
\affiliation{Universit\'e Grenoble Alpes, CNRS, IPAG, 38000 Grenoble, France}
\author{R\'{e}mi Julo}
\affiliation{Universit\'e Grenoble Alpes, CNRS, IPAG, 38000 Grenoble, France}
\affiliation{Univ. Grenoble Alpes, G. INP, CNRS, Gipsa-lab, 38400 Saint-Martin-d’H\`{e}res, France}
\author{Dorian Demars}
\affiliation{Universit\'e Grenoble Alpes, CNRS, IPAG, 38000 Grenoble, France}
\author{Sean M. Andrews}
\affiliation{Center for Astrophysics|Harvard \& Smithsonian, 60 Garden Street, Cambridge, MA 02138, United States of America}
\author{Luca Ricci}
\affiliation{Department of Physics and Astronomy, California State University Northridge, 18111 Nordhoff Street, Northridge, CA 91330,
USA}
\author{Zhaohuan Zhu}
\affiliation{Department of Physics and Astronomy, University of Nevada, Las Vegas, 4505 S. Maryland Pkwy, Las Vegas, NV, 89154, USA}
\author{Nicolas T. Kurtovic}
\affiliation{Max-Planck-Institut für Astronomie, Königstuhl 17, 69117, Heidelberg, Germany}
\author{Nicol\'as Cuello}
\affiliation{Universit\'e Grenoble Alpes, CNRS, IPAG, 38000 Grenoble, France}
\author{Xue-ning Bai}
\affiliation{Institute for Advanced Study, Tsinghua University, Beijing 100084, China}
\affiliation{Department of Astronomy, Tsinghua University, Beijing 100084, China}
\author{Til Birnstiel}
\affiliation{University Observatory, Faculty of Physics, Ludwig-Maximilians-
Universität München, Scheinerstr. 1, 81679 Munich, Germany}
\affiliation{Exzellenzcluster ORIGINS, Boltzmannstr. 2, D-85748 Garching,
Germany}
\author{Cornellis Dullemond}
\affiliation{Institute of Theoretical Astrophysics, Center for Astronomy (ZAH), Ruprecht-Karls-Universität Heidelberg, Albert-Ueberle-Str. 2,
69120 Heidelberg, Germany}
\author{Viviana V. Guzm\'an}
\affiliation{Instituto de Astrof\'isica, Pontificia Universidad Cat\'olica de Chile, Av. Vicuña Mackenna 4860, 7820436 Macul, Santiago, Chile}



\begin{abstract} 
The accretion/ejection processes in T-Tauri stars are fundamental to their physical evolution, while also impacting the properties and evolution of the circumstellar material at a time when planet formation takes place. To this date, characterization of ongoing accretion processes in stellar pairs at 5-50\,au scales has been challenging, high angular resolution spectrographs are required to extract the spectral features of each component. 
We present the analysis of spectroscopic observations of the tight (160mas, 25au) T-Tauri system HT Lup A/B, obtained with  MUSE at VLT in March and July of 2021. We focus on constraining the accretion/ejection processes and variability of the secondary component HT Lup B, by searching for accretion tracers applying High-Resolution Spectral Differential Imaging techniques. We retrieve strong (SNR $>$ 5) $H\alpha, H\beta$ and [OI]$\lambda6300$ emission in both epochs. The $H\alpha$ and $H\beta$ line fluxes showcase high variability, with variations up to 400-500\% between epochs. The fluxes are consistent with accretion rates of $8\times10^{-9} M_\odot \, yr^{-1}$ and $2\times10^{-9} M_\odot \, yr^{-1}$ for the first and second epoch, respectively. We attribute the increased accretion activity during the first night to a "burst" like event, followed by a relaxation period more representative of the common accretion activity of the system. The [OI]$\lambda6300$ line profiles remain relatively similar between epochs and suggest ejection rates on the order of $10^{-9}-10^{-10} M_\odot \, yr^{-1}$, compatible with moderate disk winds emission. Our results also indicate that the accretion processes of HT Lup B are compatible with Classical T Tauri Stars, unlike previous classifications.

\vspace{1cm}
\end{abstract}



\section{Introduction} \label{sec:intro}
  A large fraction of stars are born as binary stars or multiple systems \citep{2010ApJS..190....1R, 2017ApJS..230...15M}. Two main paradigms are used to describe their formation: turbulent fragmentation of molecular clouds and gravitational instability within a circumstellar disk of gas and dust \citep[see, e.g., ][for a review]{2022arXiv220310066O}, each one predicting different configurations and properties for the resulting systems. As stellar evolution of the components takes place, the material (dust and gas) present on the circumbinary disk and/or on individual circumstellar disks will evolve as well; but contrary to what occurs in single-star systems, their evolution will be dominated by interactions within the multiple stellar system, and thus, will be heavily dependent on the binary/multiple properties such as distance, mass ratio or eccentricity \citep{2020MNRAS.498.2936H,2020MNRAS.499.3362R}. Since these disks are the main reservoir of material for the formation of potential exoplanets, the question arises on how much the early interactions between the multiple stellar and/or sub-stellar components with the disks would impact the planet formation process.

  There is presently more than 200 planets\footnote{https://www.univie.ac.at/adg/schwarz/multiple.html} unearthed in multiple systems orbiting one component or both
components \citep[S-type and P-type, respectively;][]{1982OAWMN.191..423D}, but accessing the initial configuration of these systems to understand their formation remains challenging. Observations of pre-main-sequence binaries still in the process of formation can shed light onto this subject, as they can provide clues on the system formation process, their early properties and dynamical interactions (three-body interactions, radial migrations) between the stars, and their impact on the birthplace of planets, e.g. circumstellar and circumbinary disks \citep{1997MNRAS.285...33B, 2022arXiv221100028L}.  Tight spectroscopic T-Tauri star binaries ($<$5au) are shown to exhibit quasi periodic  “pulsed accretion” \citep[e.g.,][]{ 2013Natur.493..378M, 2014ApJ...792...64B, 2017ApJ...842L..12T, 2022ApJ...928...81F}, possibly arising from  the complex accretion streams delivered onto the young stars from a tidally truncated circumbinary disk, and depending on the orbital parameters of the pair of stars \citep[see][for a review]{2023ARA&A..61..517L}.
The impact of binarity on accretion processes and the circumstellar environment for pairs with intermediate separations (5--50\,au) has not been thoroughly investigated yet, because it requires resolved observations at optical and near-infrared wavelengths at a spatial resolution of 0.5" or better, at the distance of star forming regions ($>$100pc).

In the context of a dedicated VLT/MUSE follow-up of young planetary systems observed in the ALMA Large Program “Disk Substructures at High Angular Resolution Project” (DSHARP, \citealt{2018ApJ...869L..41A}), we observed the young system HT Lup with an estimated distance
of $154\pm2$\,pc in the Lupus star forming region \citep{2018A&A...616A..11G}. Near infrared imaging \citep[e.g.][]{1997ApJ...490..353G} and interferometry \citep{2015A&A...574A..41A} resolved HT Lup as a triple stellar system, with HT Lup A (K3, $1.3\pm0.2$\,M$_{\odot}$, $0.5^{+0.4}
_{-0.2}$\,Myr, \citealt{2022AJ....164..109R}) and B being closer in projected separation than HT Lup A/B from HT Lup C. The DSHARP ALMA high resolution observations from \citet{2018ApJ...869L..44K} constrained a projected separation between A and B of $161$\,mas (25\,au), and $2.82\,''$ (434 \,au) between AB and C. Additionally, they found that the protoplanetary disk in HT\,Lup\,A has symmetric spiral arms and small Keplerian deviations in its kinematics due to the B companion, while the disk around B is quite compact ($\sim5$ au). \cite{Kurtovic_2019} estimated the dynamical mass of HT\,Lup\,B to be $0.09$\,M$_{\odot}$, and further analysis of its rotation curve indicate that it may be a stellar/sub-stellar companion, with a gaseous disk surrounding it and most likely still accreting. The tertiary component, HT\,Lup\,C, has an estimated mass of $0.2\pm0.15$\,M$_{\odot}$ from photometry measurements \citep{2001A&A...376..982W}.

As the binary components HT Lup A and B still retain their own circumstellar disk, it is expected that their interaction will truncate the outer disk radius typically to $\frac{1}{3}$ to $\frac{1}{4}$ of the companion's orbital semi-major axis \citep{1994ApJ...421..651A}. 
If the accretion rate of the system is comparable to that of single stars of similar mass, which has already been observed for alike binary systems \citep{2001ApJ...556..265W, 2012A&A...540A..46D}, then the reduction of  disk size and available mass would result in short-lived disks that dissipate in $\sim 1$ Myr \citep{2012ApJ...745...19K}.
This will not only impact the subsequent evolution of each of the stellar components, but also the formation of planets. Given the low mass of the disks in HT Lup, is possible that the available material is not enough to trigger and sustain planet formation in the system. And even if planet formation is possible, then it would need to take place early in the lifetime of the circumstellar disks and progress quickly before the dust and gas reservoir is completely dispersed. Thus, the HT Lup system presents itself as an unique opportunity to study and characterize circumstellar accretion and explore its implications on planet formation in binary/triple systems.

This study presents VLT/MUSE observations of the close binary in the HT Lup system, showcasing signatures of ongoing variable accretion in HT Lup B. The system offers the opportunity to test accretion theory in multiple systems at Solar-System scales (5-50au). In the following, we present our specific study of the HT\,Lup\,AB system with a description of the observations in Section\, \ref{sec:obs}, the data reduction in Section\, \ref{sec:red}, and in Section\, \ref{sec:results} the results  that we  discuss in Section\, \ref{sec:Interpretation}.

\section{Observations} \label{sec:obs}
HT Lup  was observed on March 26 and July 22, 2021 with the Narrow-Field Mode of the Multi Unit Spectroscopic Explorer (MUSE-NFM)  located at VLT/UT4 \citep{2010SPIE.7735E..08B} under program 106.21EN. MUSE-NFM operates the 480--930\,nm integral field spectrograph (IFU) MUSE over a $7.5''\times7.5''$ field of view sampled by 25mas spaxels.  A field-splitter and a field separator divide the field of view into 24 sub-images sent to separate integral field units. Each unit uses a slicer to further split the sub-images into 48 slices, which are re-arranged as a pseudo-slit and dispersed by gratings onto $4k\times 4k$ CCD detectors. The spectral resolution ranges from 1740 at 480\,nm to 3450 at 930\,nm. MUSE-NFM  uses the GALACSI laser-tomography Adaptive Optics (AO) module as part of the AO facility (AOF) at UT4, to partly restore the angular resolution of the 8.2\,m telescope at optical wavelengths \citep{2008SPIE.7015E..24A, 2012SPIE.8447E..37S}. The AOF combines the use of a 4-laser guide star (LGS) system and a deformable secondary mirror. MUSE-NFM operates with an atmospheric dispersion compensator, which prevents the  field of view to drift with wavelength across the final reconstructed cubes. A notch filter within GALACSI removes wavelengths from 578 to 605nm in the IFU, corresponding to the Sodium laser emissions. 

HT Lup A was used as an on-axis Guide Source (NGS) to correct for tip-tilt aberrations and defocusing. Our March 26 and July 22 observations happened right after the first and second commissioning runs of the upgraded NGS sensing (IRLOS+), respectively, allowing for a better correction of fast turbulence and the observations of fainter guide stars.

The log of observation is reported in Table \ref{tab:obs} in Appendix \ref{App:A}. The sequence of July 22, 2021 is split into two $\sim$1h-long blocks, with total of three independent observation sequences of the target. We recorded 20s exposures at the start and end of each sequences. These short exposures were chosen to avoid saturation of the star even under excellent conditions. We then obtained 8$\times$90s exposures offering a better dynamic range on HT Lup B, and a final 150s exposure aiming for a deep search for accreting companions at large separation. The derotator at the Nasmyth focus was used to apply 45$^{\circ}$ and 90$^{\circ}$ rotations in-between  the 90s and 20s exposures, respectively. These angular offsets are used to filter out static artefacts (scattered light and hot pixels) at the data reduction step.


\section{Data Reduction} \label{sec:red}

\subsection{Pre-processing}
\label{subsec:preproc}
The data were initially calibrated using the ESO MUSE pipeline, version 2.8. Details regarding the different calibration steps can be found in the \textit{MUSE pipeline user manual} and \citet{2020A&A...641A..28W}. As a summary, the pipeline can be divided in two main steps; the \textit{pre-processing}, which apply the initial calibration frames on the basis of single CCDs, and the \textit{post-processing}, which carries out on-sky and wavelength/flux calibrations. The final product is a reconstructed 3D cube for each observation, with two spatial and one spectral dimension. 
\subsection{High Resolution Spectral Differential Imaging}
\label{subsec:HRSDI}

The MUSE cubes were analysed with the \textit{Toolkit for Exoplanet deTection and chaRacterization with IfS} (hereafter \texttt{TExTRIS}) \citep[][Bonnefoy et al., in prep]{2021A&A...648A..59P}. The tool was first used to identify and correct wavelength shifts affecting differently the 24 slicers. To do so, a model of the oxygen feature (0.76$\mu$m) was created using the ESO \textit{SkyCalc} tool \citep{refId0, refId1} and smoothed to the resolution of the instrument. The model was then linearly shifted in velocity from -100 to 100 km/s, with spacing of 0.5 km/s. At each step and for each individual spaxel, we calculated the Pearson cross-correlation coefficient between a subsection of the data centered at the oxygen feature and the shifted model, to determine the velocity correction to apply. After shifting each spaxel by its corresponding velocity, a cubic spline was fitted to retrieve the flux values for the original wavelength grid of the cube. We repeated this step, but now using a model built from the mean of the $\sim 1\%$ brightest spaxels, performing a cross-correlation of the same telluric features as above. This second step of correction relies on a data-based model of the telluric absorption and is therefore expected to increase the robustness of the calibration.  

MUSE's atmospheric dispersion compensator only partly corrects for the source motion within the field of view caused by refraction. 
The star's centroid experiences complex behavior with wavelength, such as sudden shifts in position or oscillations near red wavelengths, possibly caused by flux losses happening when the star crosses an inter-slitlet junction, sparse detector defects, or imprecise correction of the distortion. 

We used \texttt{TExTRIS} to measure the star position at each wavelength in the field of view and produce datacubes with the star located at a static spatial location (using a bicubic spline subpixel interpolation). For that purpose, we fit a 2D elliptical Moffat function to the wings of the low-Strehl AO-corrected PSF, selecting a stable fitting function from the \textit{MUSE Python Data Analysis Framework}\footnote{2016ascl.soft11003B}. The accurate and stable star registration is critical to ensure a low level of residuals at later steps.

\texttt{TExTRIS} was then used to remove the stellar halo, following the High Resolution Spectral Differential Imaging (HRSDI) method, similar to the one described in \cite{2019NatAs...3..749H}. A model of the star's spectrum was first built from the $\sim 1\%$ brightest spaxels. Each spaxel was divided by this model and the ratio was smoothed in wavelength by applying a Savitzky-Golay filter with a window width of 501 pixels, which was used to build an updated model for a given spaxel with a spectral slope close to the one of the spaxel. The use of a bigger window for the smoothing of the spectra, compared to the one used in similar works \citep[101 pix,][]{2019NatAs...3..749H, 2020A&A...644A.149X}, was done to avoid self subtraction of the lines induced by the low-pass filtering, which could mimic some real features such as blue-red shifted absorption similar to (inverse) P-Cygni profiles. This method removes the stellar emission, including any contribution from emission lines, while leaving the continuum-subtracted companion emission.  We tested PCA-based approaches using the spatial and/or spectral diversity as described in \cite{2019NatAs...3..749H}, to filter out stellar residuals caused by the non-perfect calibration of the cubes. However, this turns out not to bring any improvement in the final data quality.

During the analysis of the reduced cubes, we found that those from March presented a persistent wavelength shift, not related to the imperfect wavelength correction of the slicers, and thus, not captured on the initial wavelength calibration with \texttt{TExTRIS}. The shift was measured by cross-correlating specific photospheric features of the main binary component HT Lup A with the reference K3 star HD 160346, and the subsequent results for HT Lup B were corrected accordingly. 

\begin{figure*}[ht!]
\plotone{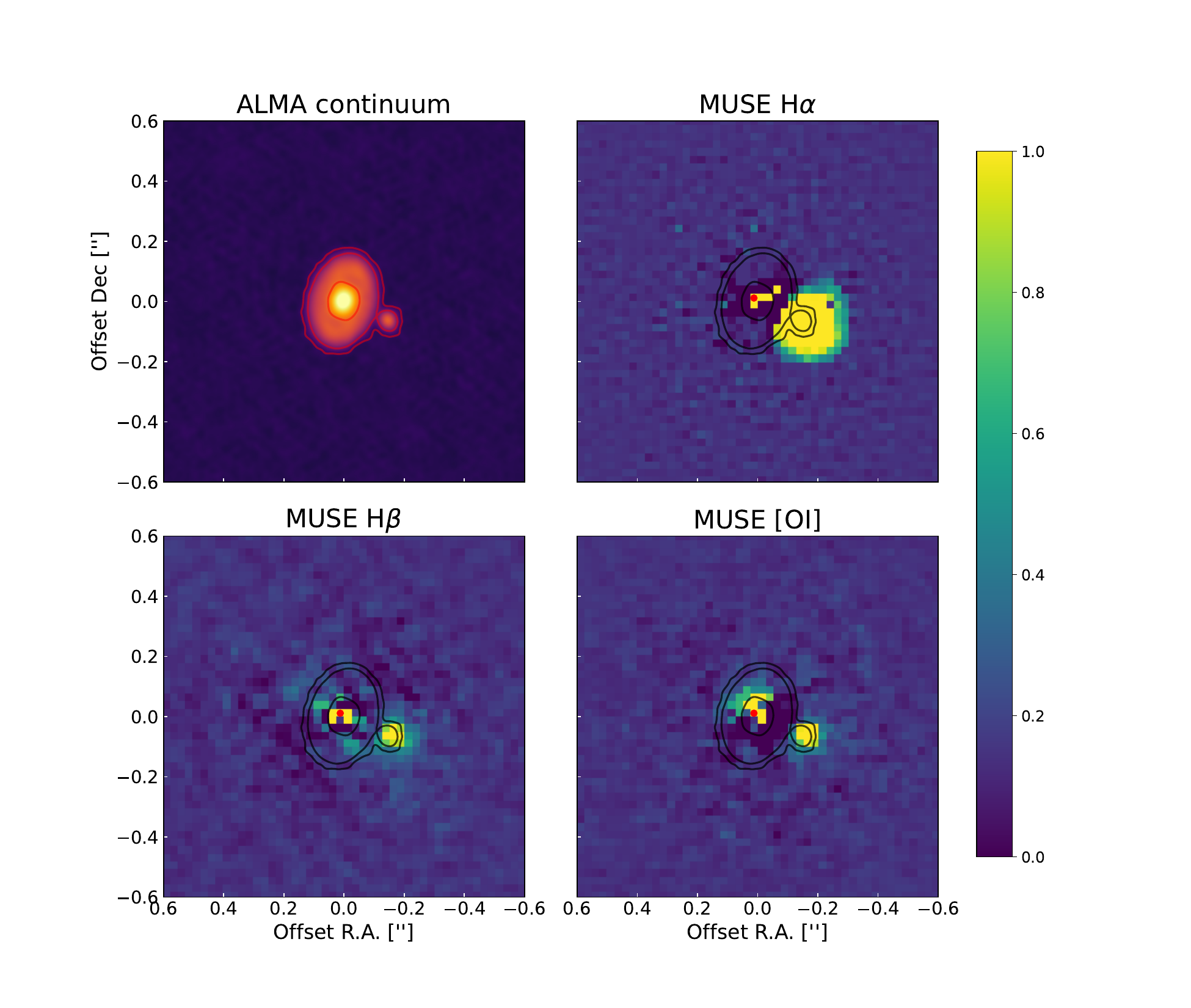}
\caption{Gallery of observations of the HT Lup system.a) ALMA 1.25 mm continuum observations of HT Lup AB from \cite{2018ApJ...869L..44K}. Contour curves correspond to the 5$\sigma$, 28$\sigma$ and 150$\sigma$ levels, showcasing the extension of the dust disk of HT Lup A as well as the location of HT Lup B. (b) H$\alpha$ (6563 \AA) image of HT Lup B after HRSDI processing. (c) H$\beta$ (4861 \AA) image of HT Lup B after HRSDI processing. (d) [OI] (6300 \AA) image of HT Lup B after HRSDI processing. All the MUSE images were equally normalized for visualization purposes.} 
\label{fig:gallery}
\end{figure*}
\section{Results}\label{sec:results}

\subsection{Detection of HT Lup B}\label{subsec:detection}

Fig. \ref{fig:gallery} showcases the result of the HRSDI reduction for HT Lup from one of our observations, together with the high-resolution ALMA observations. On all available observations we recover a clear detection of the B component at the 4851\AA, 6300\AA and 6563\AA, specifically associated with H$\beta$, [OI] and H$\alpha$ emission, respectively. 
The location of the emission is also consistent with the position of HT Lup B presented in \cite{2018ApJ...869L..44K}, although the peak location of the hydrogen emission appears slightly shifted with respect to the 1.25 mm continuum emission. Assuming this shift in position is produced by the orbital motion of HTLup B, we first proceeded to determine the relative astrometry of the B component with respect to A for both epochs. For this, we fitted a Moffat profile, which provides a close representation to the instrumental PSF, to source B at the wavelength of the peak of the H$\alpha$ emission for each independent cube, with the initial starting point determined from the brightest pixel of the source at this wavelength. The obtained position is then compared with the centroid position of HT Lup A to determine the separation between the two objects and the position angle (PA) of HT Lup B, with these results being summarized on Table \ref{tab:sep}, together with the astrometry retrieved from previous observations of the system. 
Using these results we simulated orbits using the orbitize! code \citep{2017AJ....153..229B}, which are available in Appendix \ref{App:B}. Although the orbital coverage is not enough to properly characterize the complete orbit of HT Lup B, the retrieved orbital parameters, together with the observed trend of increased separation over time while its PA remains almost constant, suggests the orbital plane of HT Lub B to be almost edge-on to the on-sky plane. It is important to point out that sub-pixel astrometry is extremely difficult with MUSE due to multiple diffraction effects caused by the slicer \citep[e.g,][]{2020A&A...644A.149X}, and thus, the obtained astrometry should be regarded as reasonable estimates rather than highly accurate.  Dedicated astrometric observations with high contrast and resolution are needed to confirm these results. Further follow up in the coming years to determine the astrometry of HT Lup B at different points in the orbit would also allow to obtain a very robust reconstruction \citep[e.g,][]{2024arXiv240402469D}

\begin{deluxetable}{ccccc}
\label{tab:sep}
\tablecaption{Relative astrometry of HT Lup B to A} 
\tablecolumns{4}
\tablehead{
\colhead{Date}  & \colhead{Instr.}  & \colhead{Sep. (mas)} &  \colhead{P.A. (deg)} &  \colhead{Ref.}}
\startdata
1997-07-01 & CTIO & $107\pm7$ & $245\pm7$ & (1)\\
2004-04-06 & VLT/NaCo & $126\pm1$ & $246.7\pm0.5$ & (2) \\
2019-05-14 & Gemini/GPI & $161\pm3$ &  $246.6\pm0.7$ & (3) \\
2021-03-25 & VLT/MUSE & $169.2\pm5$ & $250.1\pm1$ & (4) \\
2021-07-22 & VLT/MUSE & $173.6\pm4$ & $249.1\pm1$ & (4)
\enddata
\tablenotetext{}{References: (1) Ghez et al. (1997), (2) Correia et al. (2006), (3) Rich et al. (2022), (4) this work}
\end{deluxetable}

\subsection{H$\alpha$ line characterization}
\label{subsec:lineprop}
\begin{figure*}[th]
\centering
\includegraphics[width=18cm]{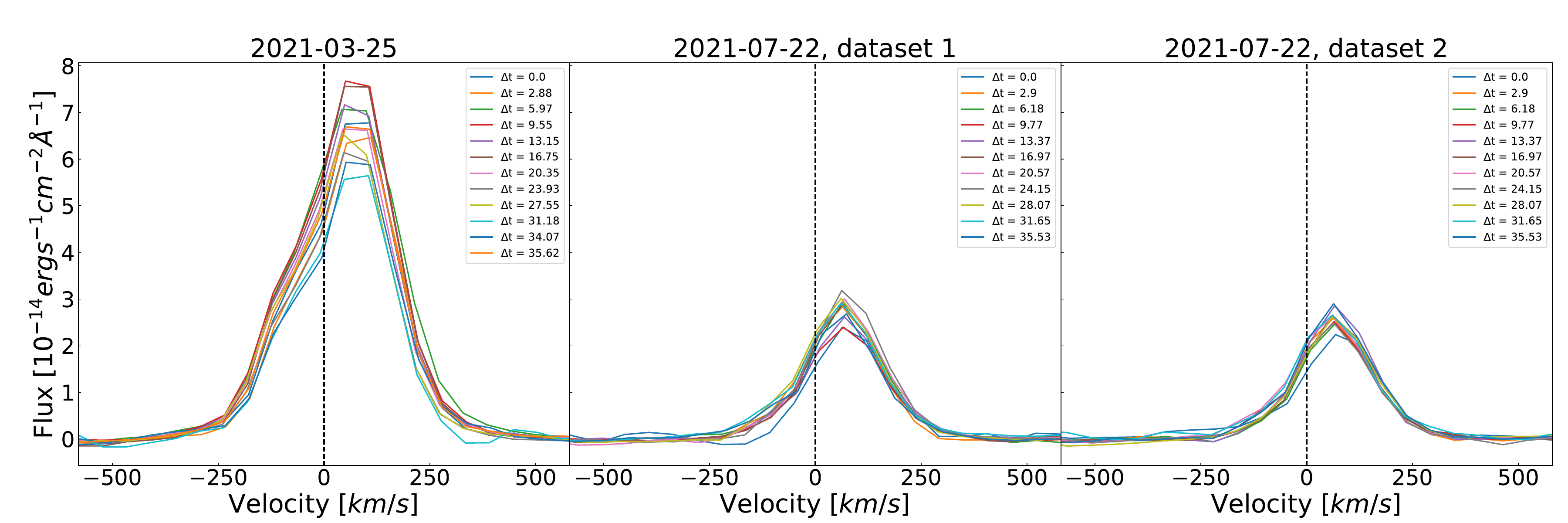}
\caption{H$\alpha$ profiles for the three  datasets of HT Lup B. Labels correspond to the time, in minutes, after the first observation of each dataset. Dashed horizontal lines mark the 0 km/s velocity, corresponding to the H$\alpha$ wavelength in vacuum (6563 \AA)}
\label{fig:line}
\end{figure*}

To obtain the $H\alpha$ line profile for each observation, we first select a subset of frames centered around the $H\alpha$ wavelength. Then, for each frame we determine the location of HT Lup B by fitting a 2D elliptic Moffat profile, identifying the location of the peak of the fitted function as the center of the emission region. We then derive the photometry at each wavelength, by using a circular aperture centered in the B companion with a radius of 3.5 pixels, which approximates the retrieved FWHM of the core PSF for the main star. We then corrected for possible flux loss due to self and over subtraction due to the HRSDI method, as well as for the impact of the small aperture used to retrieve the flux given the MUSE Strehl ratio for each observation. The details regarding each calibration is presented in Appendix \ref{App:D}. The uncertainties of the flux at each wavelength were estimated from the standard deviation of the background flux within a circular aperture with the same separation as the B companion, but with varying position angles to cover the full azimuthal extent without overlapping. The noise was then corrected for small sample statistics following \cite{2014ApJ...792...97M}. Fig. \ref{fig:line} showcases the final H$\alpha$ line profiles for the three datasets. 
To recover the total H$\alpha$ flux of the B companion at each epoch, we interpolate the line profiles by fitting a cubic spline to the data between $6550 \AA$ and $6575 \AA$, which was then integrated using the composite trapezoidal rule from the \textit{trapz} module from the numpy python package. The error was estimated using a Monte Carlo approach, by drawing N=1000 samples for each line considering the associated error at each specific wavelength and then repeating the previous integration. The final H$\alpha$ flux error corresponds to the standard deviation of the integrated flux from all samples. From the same procedure we determine important line parameters, namely the line velocity and the $10\%$ and $50\%$ linewidths ($W_{10\%}$ and $W_{50\%}$, respectively). The line velocity was determined as the velocity of the peak of the emission line, while the respective widths were calculated with respect to the flux at this point. Table \ref{tab:lines} summarizes all the aforementioned results for each individual observation.


We were able to recover a mean apparent H$\alpha$ flux of $2.1 \pm 0.1 \times 10^{-13}$ erg s$^{-1}$ cm$^{-2}$, $6.8 \pm 0.2 \times 10^{-14}$ erg s$^{-1}$ cm$^{-2}$ and $6.2 \pm 0.1 \times 10^{-14}$ erg s$^{-1}$ cm$^{-2}$ for the first, second and third set of observations, respectively. We corrected for interstellar extinction using the extinction law from \cite{2007ApJ...663..320F} with a value of $A_{V} = 1.0$ \citep{2017A&A...600A..20A}. This gives a final apparent flux of  $4.2 \pm 0.1 \times 10^{-13}$ erg s$^{-1}$ cm$^{-2}$, $1.4 \pm 0.2 \times 10^{-13}$ erg s$^{-1}$ cm$^{-2}$ and $1.3 \pm 0.1 \times 10^{-13}$ erg s$^{-1}$ cm$^{-2}$.

The mean line velocity for the three datasets are $V_1 = 75.4 \pm 3.4$ km/s, $V_2=66.1 \pm 4.1$ km/s and $V_3=64.4 \pm 4.6$ km/s, respectively, suggesting a slight velocity shift of the line between epochs. Given the spectral FWHM of the MUSE-NFM is $\sim 114$ km/s ($\sim 2.5$ \AA) at these wavelengths, we checked for possible calibration effects that could explain a line velocity shift between epochs, by retrieving and analyzing H$\alpha$ line profiles from the main binary component HT Lup A. These profiles revealed no significant shift with respect to the rest velocity at the H$\alpha$ wavelength on any of the datasets, strongly suggesting that the observed velocity shift on the HT Lup B line profiles are a real variation and not a calibration issue (see further discussion in Sect. \ref{subsec:physprop}).

\subsection{Variability}
\label{subsec:varlev}
\begin{figure}[h]
\centering
\epsscale{1.3}
\plotone{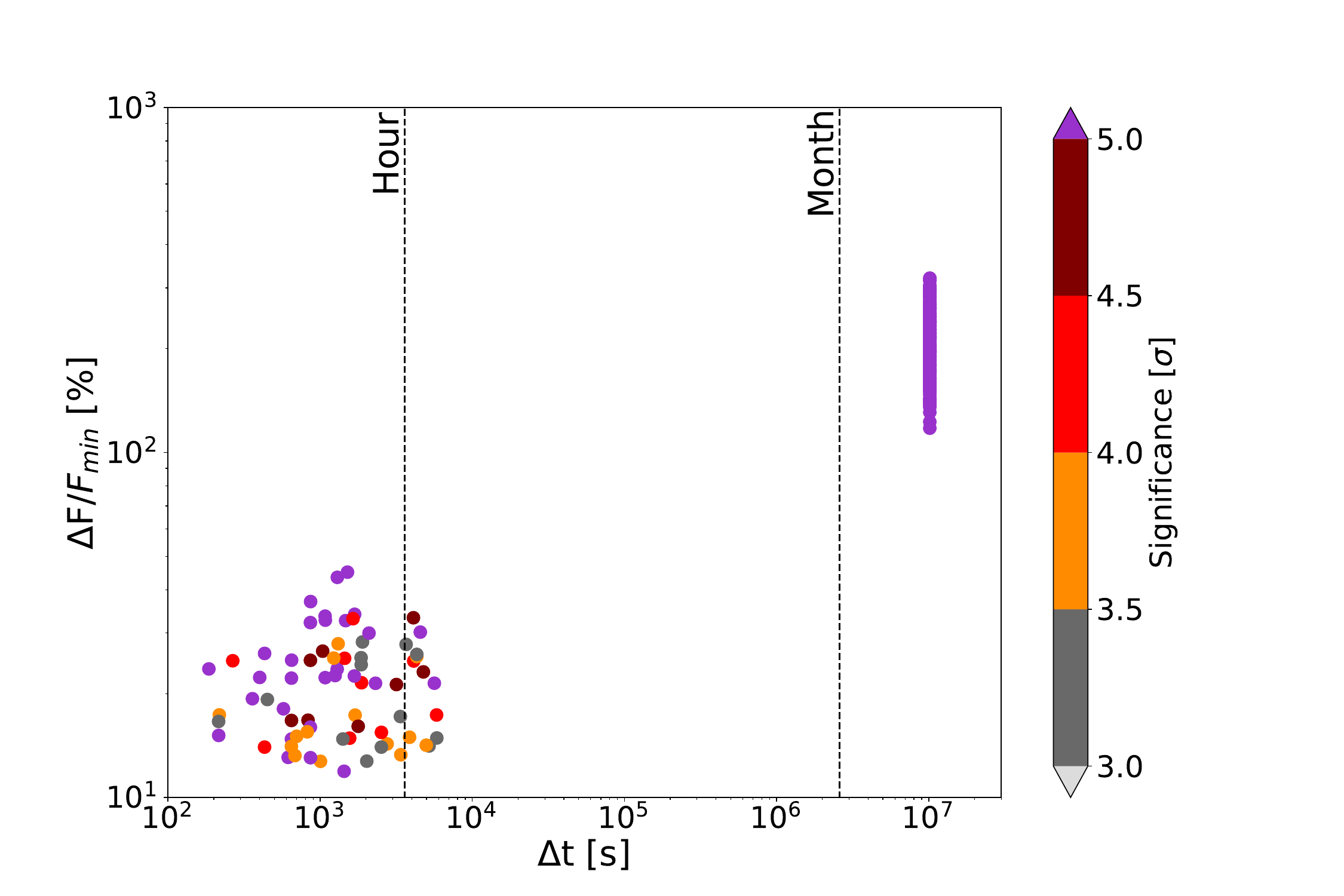}
\caption{H$\alpha$ flux variability as a function of time. Each measurement corresponds to the difference in flux for an individual pair of lines from the full set of observations, divided by the minimum flux of the pair. The color of each point represents the $\sigma$-distance between the lines used to compute the flux variation. Grey points indicate a value of $\sigma < 3$, considered to show no flux variation given the uncertainties on the flux measurement for each pair. Colored points correspond to values of $\sigma \geq 3$, indicating real flux variation.}
\label{fig:fluxvar}
\end{figure}

\begin{figure*}[th!]
\centering
\includegraphics[width=17.8cm]{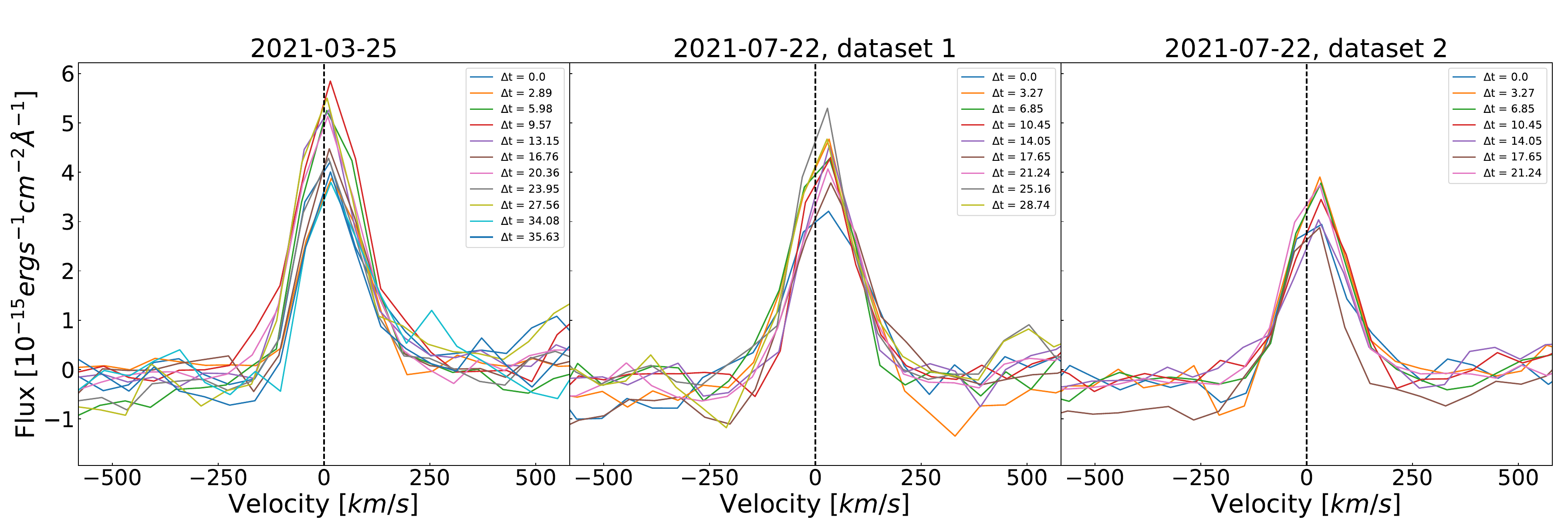}
\caption{[OI]$\lambda6300$ profiles for the three available datasets of HT Lup B. Labels correspond to the time in minutes after the first observation of each dataset. Only lines with SNR higher than 10 at the peak of the emission were included for the analysis. Dashed horizontal lines mark the 0 km/s point, corresponding to the [OI] wavelength in vacuum (6300 \AA)}
\label{fig:OIline}
\end{figure*}

As seen on Fig.\,\ref{fig:line}, H$\alpha$ emission is identifiable on all observations from the three datasets, with clear variation after the 4 months time span between the first and last set of observations. In this section we focus on quantifying the variability on the H$\alpha$ line profiles and fluxes.


Regarding the line profiles, clear differences can be observed and identified between both epochs based on the $10\%$ and $50\%$ linewidths. For the first dataset, we have a mean value of $W_{1, 10\%} = 479.2 \pm 2.5$ km/s and $W_{1, 50\%} = 250.0 \pm 5.8$ km/s, while the derived values for the other two datasets correspond to $W_{2, 10\%} = 420.7 \pm 6.3$ km/s, $W_{2, 50\%} = 194.0 \pm 4.6$ km/s and $W_{3, 10\%} = 417.1 \pm 8.6$ km/s, $W_{3, 50\%} = 191.9 \pm 5$ km/s respectively. The last two datasets present good agreement between their line profiles, as expected given to the fact that they were obtained during the same night. The profiles retrieved during the first epoch are noticeably wider, possibly related to different accretion events taking place at each epoch. Specifically, the H$\alpha$ line appears to narrow by $\sim 60$ km/s after the 4 months between observations.

We also study the variability of the H$\alpha$ flux in short and long timescales. We compute the relative flux variation between a pair of lines as $(F_{max}-F_{min})/F_{min}$, and register the $\sigma$-distance between the compared lines as a measurement of the statistical significance of the relative flux variation. This comparison is repeated for all possible combinations of two lines from our sample, to assess the consistency of the registered variability as a function of time. Fig. \ref{fig:fluxvar} summarizes our results.
We consider cases with $\sigma < 3$ to be compatible with no flux variation, as this indicates that the flux level of the pair of compared lines are not significantly different, given our uncertainties. From the cases of actual flux variation ($\sigma \geq 3$), we register two distinct cases; minutes (short) timescale variability around $20\%-30\%$ level, and month (long) variability showing $200\%-300\%$ flux variation, compatible with what is shown on Fig. \ref{fig:line}. We point out that the registered short time variability comes mainly from the first epoch (2021-03-26) observations, as the second and third datasets show relatively uniform H$\alpha$ emission. 

Both the line morphology and flux variations point to different physical processes taking place in the HT Lup B system at different epochs. A detailed analysis on the origin of the observed variability is discussed in Sect. \ref{subsec:varorig}

\subsection{Forbidden [OI] line detection}

\label{subsec:OIline}
\begin{deluxetable*}{ccccccc}
\label{tab:OIfit}
\tablecaption{Gaussian fitting parameters for [OI]$\lambda$6300} 
\tablecolumns{7}
\tablehead{\colhead{Date} & \colhead{$\mu_{\text{LVC}}$}  &\colhead{$\sigma_{\text{LVC}}$}  &\colhead{FWHM\textsubscript{int,1}} & \colhead{$\mu_{\text{HVC}}$} & \colhead{$\sigma_{\text{HVC}}$} & \colhead{FWHM\textsubscript{int,2}} \\
& ($km \: s^{-1}$) & ($km \: s^{-1}$) & ($km \: s^{-1}$) & ($km \: s^{-1}$) & ($km \: s^{-1}$) & ($km \: s^{-1}$) }
\startdata
2021-03-25 & $5.9\pm4.5$ & $61.8\pm3.1$ & $83.6$  & $106.7\pm34$ & $90.7\pm14.1$ & 177.3\\
2021-07-22 & $26.4\pm1.1$ & $68.6\pm1.2$& $109.4$&-& -&-\\
2021-07-22 & $24.4\pm1$ &$64\pm0.9$& $92.7$ &-& -&-\\
\enddata
\tablenotetext{}{Best-fit parameters of the gaussian fit to the [OI]$\lambda$6300 emission line for each dataset. The FWHM\textsubscript{int} corresponds to the intrinsic width of the line, computed as the square root of the quadratic difference between the FWHM of the retrieved lines and the LSF FWHM of MUSE. For the first epoch, the best fit consists of a double gaussian profile with both components being independently resolvable based on their retrieved widths. For both datasets of the second epoch, the best fit reduces to a single gaussian profile. LVC and HVC refers to a low and high velocity component, respectively, as defined in Sect. \ref{subsec:ejecrate}}
\end{deluxetable*}

\begin{figure*}
\centering
\includegraphics[width=17.8cm]{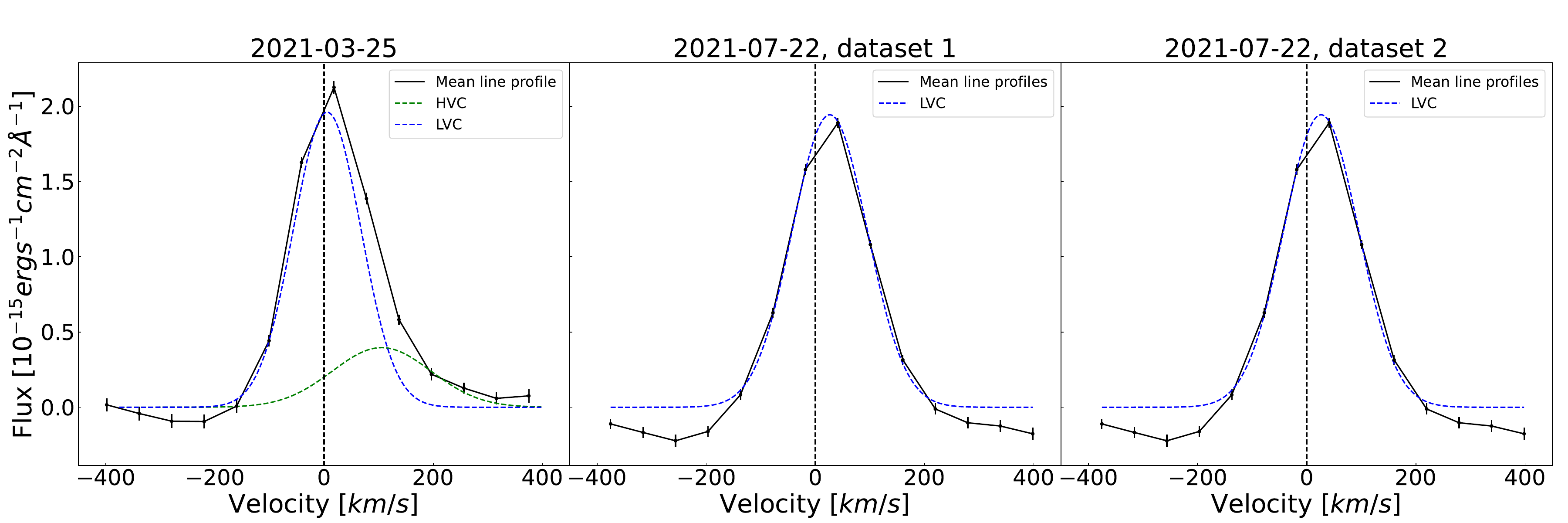}
\caption{[OI]$\lambda6300$ line profile fits for the three available datasets of HT Lup B. Black lines correspond to the mean line profile from each dataset. For the first epoch, the best fit consists of a double gaussian profile, while both datasets of the second epoch are better fitted by a single component. Green and blue dashed lines represent the HVC and LVC respctively, as defined in Sect.\ref{subsec:ejecrate} Dashed horizontal lines mark the 0 km/s point.}
\label{fig:OIdecomp}
\end{figure*}

 We were able to retrieve clear [OI]$\lambda6300$ lines for HT Lup B, presented in Fig. \ref{fig:OIline}. The same methodology described in section \ref{subsec:lineprop} was used to retrieve the [OI]$\lambda6300$ emission. Although we were able to obtain line profiles for all observations, we only include lines with $SNR\geq10$ for subsequent analysis, as [OI]$\lambda6300$ is intrinsically fainter than the H$\alpha$ case, and more impacted by the residual noise. We were able to retrieve a mean apparent [OI]$\lambda$6300 flux of $1.8 \pm 0.1 \times 10^{-14}$ erg s$^{-1}$ cm$^{-2}$, $1.5 \pm 0.1 \times 10^{-14}$ erg s$^{-1}$ cm$^{-2}$ and $1.1 \pm 0.1 \times 10^{-14}$ erg s$^{-1}$ cm$^{-2}$ for the first, second and third dataset respectively, after correcting for extinction.

To further characterize the [OI]$\lambda$6300 emission of the system, we fitted the mean emission line of each dataset following a standard Gaussian fitting approach \citep[e.g,][]{2016ApJ...831..169S}, using the Markov-Chain Montecarlo method with the EMCEE python package \citep{2013PASP..125..306F} to retrieve the posterior distribution of the line parameters, namely the amplitude $A$, peak velocity $\mu$ and width $\sigma$. We consider a set of uniform priors for all parameters, with the restriction that $A>0$.  As a first step, we determine if the retrieved lines are either fully resolved, marginally resolved, or unresolved, by comparing the retrieved width of a single Gaussian fit with the resolution limit of MUSE (2.5\AA\,  or 119 km/s at 6300\AA). For all cases we retrieved a $\text{FWHM}>150$ km/s, indicating that the line was fully resolved at all epochs. As the [OI]$\lambda6300$ line can be produced as a combination of multiple components, associated with emission from different sources or regions of the disk, we repeated the process considering 2 and 3 Gaussian components, computing the RMS as a measure of the goodness of each fit. For the first epoch, the line was better reproduced considering a double gaussian model, while a single gaussian model better reproduces the second epoch line profile. The parameters of the best fit for each dataset are presented in Table \ref{tab:OIfit}, while the respective line decomposition are showcased inf Fig. \ref{fig:OIdecomp}. For both epochs, the [OI]$\lambda$6300 lines present a clearly different shift than their respective H$\alpha$ counterparts. A possible explanation for this discrepancy is the fact that these lines trace different regions of the system; while H$\alpha$ emission is mostly produced at the accretion funnel and shock regions on the star, [OI]$\lambda$6300 is a tracer of outflows (jets/winds) arising from the disk, and could have a different projected velocity depending on the orientation of the outflow itself and the region of the disk where it originates (Further discussion in Sect. \ref{subsec:ejecrate}).







\begin{figure*}[ht!]
\centering
\includegraphics[width=17.8cm]{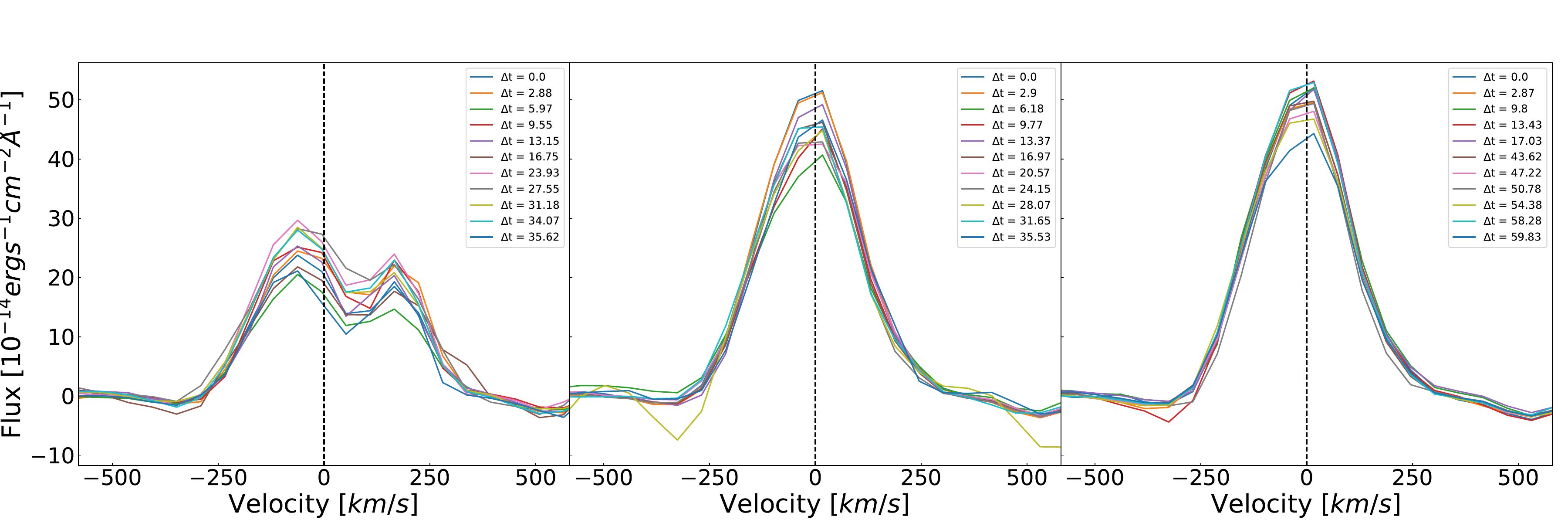}
\caption{H$\alpha$ profiles for the three available datasets of HT Lup A. Labels correspond to the time in minutes after the first observation of each dataset. Dashed horizontal lines mark the 0 km/s point, corresponding to the H$\alpha$ wavelength in vacuum (6563 \AA). Besides the higher H$\alpha$ flux level on HT Lup A, the lines exhibit clear distinction with those from the secondary. In particular, the emission lines at the first epoch showcase an slightly redshifted absorption feature, not found on any other observations. The data was carefully inspected to discard any artificial origin for this feature, such as saturation of the data.} The variation of the line flux follows an opposite trend to that of HT Lup B; while the lower flux in the second epoch appears to indicate a more relaxed, quiescent state for HT Lup B, for HT Lup A emission increases compared to the first, possibly indicating different accretion events taking place in HT Lup A and B.
\label{fig:HTLupA}
\end{figure*}

\subsection{H$\alpha$ line comparison with HT Lup A}
\label{subsec:comp}

 In this section we focus on the primary star HT Lup A, as it is a crucial step to understand the interactions taking place in the system and how they influence their subsequent evolution. 

First, we retrieve the spectra of HT Lup A on all observations to check for the presence of any emission lines indicative of ongoing accretion. The procedure, final spectra and properties of the detected emission lines are presented in detail in Appendix \ref{App:E}, while the HT Lup A H$\alpha$ lines are presented in Fig. \ref{fig:HTLupA}.All observations reveal strong H$\alpha$ line emission, and clearly differ with the ones obtained for HT Lup B at each epoch, as the  HT Lup A lines are generally wider with a mean 10\% and 50\% linewidth of $W_{1, 10\%} = 545 \pm 34.7$ km/s $W_{1, 50\%} = 402.6 \pm 7.4$, $W_{2, 10\%} = 468.7 \pm 18.4$ km/s, $W_{2, 50\%} = 263.5 \pm 4.5$ km/s and $W_{3, 10\%} = 465.7 \pm 10.4$ km/s, $W_{3, 50\%} = 262.8 \pm 3$ for the first, second and third dataset respectively. This translates to a 50\% linewidth difference of $\sim 154$ km/s for the first epoch and $\sim 72$ km/s for the second epoch, clearly above our uncertainties for both cases and indicating a real difference between the lines, rather than a product of the data processing.  At the same time, the velocity shifts of the lines are notably different for each target. While the HT Lup B lines showcase redshifted emission for all epochs on the order of 70 km/s, the lines from HT Lup A are much closer to rest velocity, with $V_1 = 21.9 \pm 9.6$ km/s, $V_2=2.8 \pm 5.2$ km/s and $V_3=2.3 \pm 3.9$ for each  dataset. Finally, the morphology of the HT Lup A H$\alpha$ line during the first epoch is clearly distinct, not only from HT Lup B, but also from the same target at different epochs, with a strong absorption feature being present during the first night. All of these features further support our detections to be from emission coming from HT Lup B, as the differences in the lines are likely related to distinct local processes on each source.

\section{Interpretation}
\label{sec:Interpretation}

\subsection{Stellar properties from $H_{\alpha}$ emission}
\label{subsec:physprop}

Generally, T Tauri stars are divided into Classical T Tauri stars (CTTS), which normally exhibit a circumstellar disk and H$\alpha$ line emission associated of strong accretion processes, and Weak-line T Tauri stars (WTTS), with either a very depleted disk or no disk at all and with H$\alpha$ line emission being mostly a result of chromospheric activity. A standard approach to classify T Tauri objects as either CTTS or WTTS is to measure the equivalent width ($EW$) of the $H_{\alpha}$ line, with an $EW \geq 10$\AA being associated with CTTS, as $H_{\alpha}$ emission from chromospheric activity rarely surpasses this value. Under this assumption, the HT Lup A/B system has been originally classified as a WTTS \citep{1994AJ....108.1071H, 2005A&A...443..541G}, although it was not possible to resolve each component independently. However, high-resolution spectroscopy observations on T Tauri objects have found that the $H_{\alpha}$ line width measured at $10\%$ of the maximum and its morphology a more sensitive diagnostic for accretion activity, with actively accreting (CTTS) objects normally presenting broad lines with $W_{10\%} \geq 270 \, \textrm{km s}^{-1}$ and either symmetric or asymmetric profiles \citep{2003ApJ...582.1109W, 2004A&A...424..603N}. Under this prescription, HT Lup A showcase an H$\alpha$ profile more akin to CTTSs \citep{2013ApJ...762..100C}, and thus its classification has remained ambiguous. As our observations allowed us to resolve the H$\alpha$ profiles for both HT Lup A and B, we are capable to classify both sources separately for the first time. For both stars, the H$\alpha$ linewidths are clearly above the $270 \, \textrm{km s}^{-1}$ threshold on all datasets, based on the values presented in sections \ref{subsec:varlev} and \ref{subsec:comp}. This, combined with the detection of other emission lines linked with accretion processes and the presence of circumstellar disks as revealed by ALMA, strongly suggest that both HT Lup A and B are better correlated to CTTSs, rather than WTTSs. The subsequent analysis thus assumes this classification, and relates the observed line emission to different accretion processes.

For accreting pre-main sequence stars, the $H_{\alpha}$ line flux and morphology have been correlated to different stellar and accretion properties, such as accretion luminosity ($L_{acc}$) and accretion rate, and by extension to the mass of the stellar object. Here, we derive the properties of HT Lup B separately, using the recovered values of $H_{\alpha}\, 10\%$ widths and the total $H_{\alpha}$ line flux to obtain a more comprehensive characterization of the system. Due to the observed variability of the $H_{\alpha}$ line, these properties will be derived separately for each epoch, and should be considered as relative limits rather than precise retrievals.

\cite{2004A&A...424..603N} found an empirical relation between $H_{\alpha}\, 10\%$ width and the accretion rate for low mass objects. Although the spread of this relation has been found to be rather large \citep[e.g,][]{2008ApJ...681..594H} due to additional effects that could alter the line profile such as the inclination of the system \citep[e.g,][]{2006MNRAS.370..580K} or rotational broadening \citep[e.g,][]{2022MNRAS.514.2162W}, it has the advantage of being independent to the stellar extinction and veiling effects, that are not well constrained for this particular system. We derive a mean accretion rate of $5.5 \times 10^{-9} M_\odot \, yr^{-1}$, $1.5 \times 10^{-9} M_\odot \, yr^{-1}$  and $1.2 \times 10^{-9} M_\odot \, yr^{-1}$ for the first, second and third datasets, respectively. These results indicate a strong level of accretion for HT Lup B, up to one order of magnitude higher compared with objects of similar mass in the Lupus region \citep{2017A&A...600A..20A}. However, the derived accretion rates are still in relatively good agreement with the overall accretion rates in Lupus, as expected given the known correlation between accretion rate and age for this type of systems \citep{2000prpl.conf..377C, 2014A&A...572A..62A}.


Alternatively, we can use the total $H_{\alpha}$ luminosity to characterize HT Lup B. Although it is not the best accretion line for this type of characterization, as being more affected by veiling and chromospheric emission, it still allows for a better determination on the accretion properties of the system since the correlation between accretion luminosity and accretion rate is better constrained than for the $H_{\alpha}\, 10\%$ width case. We first compute the $H_{\alpha}$ line luminosity as $L_{H\alpha} = 4\pi d^2 f_{H\alpha}$, with $d$ being the distance to HT Lup A and assumed to be the same for HT Lup B, and $f_{H\alpha}$ being the extinction corrected flux of the line. We then use the relation from \cite{2017A&A...600A..20A}, to convert the $H_{\alpha}$ line luminosity to the total accretion luminosity of the system.


Finally, the accretion luminosity can be used to calculate the accretion rate of the system using the relation:
\begin{equation}
    \dot{M}_{\textrm{acc}} = \left(1 - \frac{R_{\star}}{R_{\textrm{in}}}\right) \frac{L_{\textrm{acc}}R_{\star}}{GM_{\star}} \approx 1.25\frac{L_{\textrm{acc}}R_{\star}}{GM_{\star}}
\end{equation}
Where $R_{\star}$ and $M_{\star}$ are the HT Lup B radius and mass, respectively, and $R_{\textrm{in}}$ is the inner disk radius, which was assumed to be $3R_{\star}$ from previous studies. We use the 0.09 $M_{\odot}$ estimated mass by \cite{Kurtovic_2019}, as no other mass estimates are available. The stellar radius was considered to be 1.11$R_{\odot}$ from the \cite{2015A&A...577A..42B} evolutionary models, based on the mass and age of the system.

Using this aproach, we derive a mean accretion rate of $2.9 \times 10^{-9} M_\odot \, yr^{-1}$, $8.2 \times 10^{-10} M_\odot \, yr^{-1}$  and $7.4 \times 10^{-10} M_\odot \, yr^{-1}$ for the first, second and third datasets, respectively. These results are  one order of magnitude lower than the ones derived from the  $H_{\alpha}\, 10\%$ linewidth, and are in better agreement with other similar objects from Lupus.

Regarding the redshift of the $H_{\alpha}$ lines at all epochs, it is not straightforward to link it with physical and/or orbital properties of the system. It has been found that some T Tauri stars exhibit some naturally redshifted $H_{\alpha}$ emission, on the order of 10-20 km/s \citep{2000AJ....119.1881A}, which could partially explain the observations. Modelling of emission due to accretion processes have also found a correlation between the morphology of different lines with the inclination of the star. In the particular case of $H_{\alpha}$, at inclinations $i \geq 38^{\circ}$ the line starts to exhibit apparent blueshifted absorption, due to the presence of stellar/disk winds, with the strength of the absorption feature increasing with inclination \citep{2011MNRAS.416.2623K, 2012MNRAS.426.2901K}. However, the specific morphology of the absorption is heavily dependant on both the accretion and wind parameters, which are not completely determined for HT Lup B (see Section \ref{subsec:ejecrate} for details regarding characterization of winds on the system). 
We finally compare the accretion events ocurring in HT Lup A and B, to check for any distinct features between the two sources. We first derive the accretion rate of HT Lup A from its H$\alpha$ luminosity, following the same procedure as for HT Lup B in Sect. \ref{subsec:physprop}. We opt for this approach rather than using the 10\%  width of the line, since we have better constrains on the stellar parameters of the primary, and to avoid the higher dispersion relation of linewidth with accretion rate. We derive a mean accretion rate of $\dot{M}_{\textrm{acc},A,1} = 4.3 \times 10^{-9} M_\odot \, yr^{-1}$, $\dot{M}_{\textrm{acc},A,2} = 5.2 \times 10^{-9} M_\odot \, yr^{-1}$  and $\dot{M}_{\textrm{acc},A,3} = 5.8 \times 10^{-9} M_\odot \, yr^{-1}$ for the first, second and third datasets, respectively. For all observations, the derived accretion rate of HT Lup A is noticeable higher than HT Lup B, being almost double during the first epoch and eight times higher at the end of the second epoch. Such accretion rates are in good agreement with individual sources in the Lupus region of similar mass \citep{2014A&A...561A...2A, 2017A&A...600A..20A} and are consistent with individual circumstellar accretion, with the accretion rate increasing with the mass of the object \citep{2012A&A...540A..46D,2017A&A...600A..20A}.

\subsection{Possible origins of H$\alpha$ line variability}
\label{subsec:varorig}
As showcased in section \ref{subsec:varlev}, the $H_{\alpha}$ line profiles of HT Lup B exhibits variability from minutes to months timescales, both in their flux level as well as their morphology. Exact determination of the mechanisms that originate such variations remains a difficult task, not only because of the lack of coverage on days to weeks timescales, already been associated to fluctuations mostly due to rotational modulations \citep{2015A&A...581A..66V, 2017MNRAS.465.3889R}, but also due to the different processes to which CTTS are subjected that could contribute to its variability. Nonetheless, the marked differences on the line profiles between the two epochs, allow for at least a preliminary determination of the origin of the observed variability.

Based on the $W_{10\%}$ criteria discussed in the previous subsection, we consider that the overall $H_{\alpha}$ emission is mostly a product of active accretion on both epochs, rather than chromospheric or wind variability. It is also unlikely that the observed fluctuations come from stellar flare activity, as their $H_{\alpha}$ emission is normally concentrated in the line centre, with wings extending <100 km/s \citep{1990ApJS...74..891R}, and showing exponential decay not consistent with our observed variations \citep[e.g,][]{2010A&A...514A..94L}. Consequently, we expect the variations on the line profiles to be related to variations on accretion processes instead. Specifically for CTTSs, for which magnetospheric accretion is the most commonly associated accretion scenario, such changes in accretion are mostly linked to fluctuations on the amount of mass transferred from the disk to the star, via  accretion funnels following the magnetic field lines of the star. 

From the results of section \ref{subsec:physprop}, the variability of the mean accretion rate is $\sim 200\%$ between the two epochs. Based on both the level and timescale of these fluctuations, we are likely observing an accretion burst episode occurring on the first night, which can have a duration up to a few days \citep[e.g,][]{2012A&A...541A.116A, 2014AJ....147...83S}, followed by a more quiescent accretion phase captured on the second epoch. The origin of such burst events is not completely clear, but it appears to be linked to unstable accretion processes and instabilities on the  inner disk \citep[e.g,][]{2016AJ....151...60S, 2020AJ....160..221C}; in the case of binaries, such events can also be triggered due to star-disk interaction between the components \citep{2017ApJ...835....8T}. This scenario can also tentatively explain the shorter timespan variations observed on the first epoch. Instabilities on the disk or the magnetic field, specially in cases with strong accretion rates, would produce a "clumpy", inhomogeneous flow of material on the accretion funnels and towards the stellar surface, and as the timescale for the infall of gas into the star less than 1 hour \citep{1994A&A...287..131G}, this will inevitably produce short lived variability on the scale of a few minutes. 

We compare our results with variability studies reported for other T Tauri systems, to check for any discrepancies with more typical variability regimes. \cite{2022A&A...660A.108Z} reported $H_{\alpha}$ variability  for the system CR Cha spanning a wide range of timescales. Their results are in very good agreement with what we observe for HT Lup B, showcasing fluctuations on both short and long timescales, with accretion variability having a maximum of intensity on weeks to months timescales, as seen for HT Lup B. Their observed line morphology variations are also consistent with our observations, with short, minute variations having similar line shapes and only changing on the overall flux level of the line, while longer, monthly to yearly variations, having distinct differences between the line profiles. Our results are also in line with the reports from \cite{2014MNRAS.440.3444C}, which found that accretion variability is mostly dominated by slow, gradual variations over timespans of days to weeks after strong accretion periods, while still exhibiting rapid events on timescales of minutes. On both studies, variability was mostly attributed to rotational modulation of the accretion rate, in which different regions of the accretion flow become visible as the star rotates. This translates to variations of the derived accretion rates of around $200\%$. This however is not enough to explain the differences retrieved for HT Lup B, which exhibits a difference of $\sim 350\%$ based on the results of Section \ref{subsec:physprop}. This supports the accretion burst explanation proposed for HT Lup B, as additional processes need to be considered to account for the strong variability between the two epochs.

It is important to point out that although HT Lup B showcases a relatively common accretion pattern, and that the proposed burst scenario could very well explain the strong variation between the first and second epoch of observations, our datasets are still insufficient to provide a complete picture of the accretion regime of the system. 
Particularly, we lack coverage of the day to week timescales, which could possibly reveal a more periodic varaiability pattern possibly related to rotational modulation, and the binary period and eccentricity \citep{2022MNRAS.514..906C, 2022arXiv221100028L}.  Further observations, together with dedicated models for the HT Lup system, are necessary to provide a more complete characterization of its accretion history.

\subsection{H$\beta$ emission on HT Lup B}
\label{subsec:hbeta}

\begin{figure}[h!]
\centering
\epsscale{1.2}
\plotone{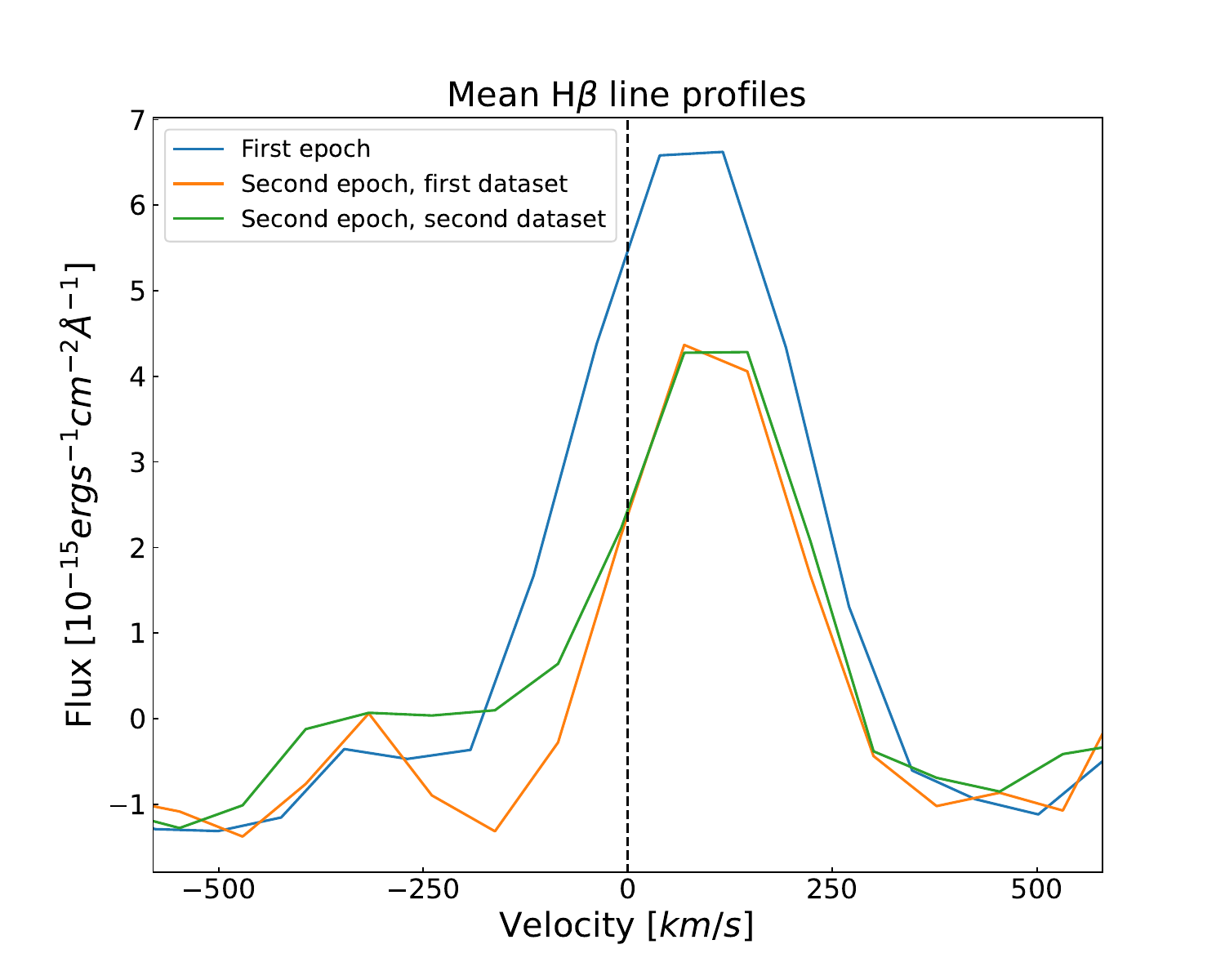}
\caption{Average H$\beta$ line profiles in HT Lup B for each dataset. Each profile was computed considering only observations with SNR>5 for the H$\beta$ line}
\label{fig:hbeta}
\end{figure}

Besides the main H$\alpha$ detection, we were also able to retrieve emission from an additional hydrogen transition line, H$\beta$, as presented in Figure \ref{fig:gallery}. However, as these lines are considerably weaker than their respective H$\alpha$ counterparts and detection is heavily limited by noise, we reduce the sample so to only consider the cases where SNR$> 5$. Figure \ref{fig:hbeta} showcases the average profile for each datasets.

As a first step, we characterize the H$\beta$ lines following the same procedure described in \ref{subsec:lineprop}.   We were able to recover a mean total flux of $1.1 \pm 0.1 \times 10^{-14}$ erg s$^{-1}$ cm$^{-2}$, $5.5 \pm 0.6 \times 10^{-15}$ erg s$^{-1}$ cm$^{-2}$ and $6.1 \pm 1.2 \times 10^{-15}$ erg s$^{-1}$ cm$^{-2}$ for the first, second and third datasets, respectively. Correcting for interstellar extinction gives a respective final flux of  $3.2 \pm 0.2 \times 10^{-14}$ erg s$^{-1}$ cm$^{-2}$, $1.6 \pm 0.2 \times 10^{-14}$ erg s$^{-1}$ cm$^{-2}$ and $1.8 \pm 0.4 \times 10^{-14}$ erg s$^{-1}$ cm$^{-2}$. Regarding the widths of the lines, we retrieve values of $W_{1, 10\%} = 411.8 \pm 30$ km/s and $W_{1, 50\%} = 245.2 \pm 19$ km/s, $W_{2, 10\%} = 288 \pm 41.2$ km/s, $W_{2, 50\%} = 176.9 \pm 26.3$ km/s and $W_{3, 10\%} = 313.4 \pm 65.8$ km/s, $W_{3, 50\%} = 183 \pm 34.5$ km/s, for each dataset respectively.

While is not possible to do a similar H$\beta$ variability analysis as the one presented in Section \ref{subsec:varlev}, due to the reduced size of the sample, we can still study the clear variation between epochs to see how it compares to H$\alpha$. Regarding the total flux of the line, in this case it is reduced by a factor of $\sim1.9$ in the 4 months period. Although the variation was slightly stronger for H$\alpha$ ($\sim3$), the results are still in good agreement for both hydrogen lines. As H$\beta$ emission is also an expected product of accretion processes, the consistent variation between both lines further supports that the differences in emission are produced by changes on the accretion regime, and that the enhanced emission on the first epoch is a product of an accretion burst event.

Additional evidence that supports the active accretion scenario comes from comparing the H$\alpha$ and H$\beta$ linewidths at each respective epoch. \cite{2017A&A...599A.105A} found from a sample of actively accreting T Tauri objects in Lupus that the emission lines normally fall into one of two groups, either narrow symmetrical lines with FWHM below 100 km/s, or wide lines with FWHM around 200-250 km/s. Our results clearly indicate that HT Lup B would be classified as part of the second group at either epoch. The same study found that for this group the H$\alpha$ is either wider or has similar width that H$\beta$. This is also in good agreement with the results obtained here for HT Lup B.

Analysis of the line ratios of the different hydrogen transitions from the Balmer series, commonly known as the Balmer decrements, has already been proved as an effective approach to study the regions from where emission lines are produced, and to derive properties such as temperature and gas density \citep{2015A&A...575A...4F,2014A&A...570A..59W,2017A&A...599A.105A}. Proper analysis requires retrieval of multiple Balmer decrements, but it is still possible to provide a preliminary characterization using our available H$\alpha$ to H$\beta$ ratios ($F_{H\alpha}/F_{H\beta}$). 


We also explore an alternative scenario from \cite{2011MNRAS.411.2383K}, which considers local line excitation and solves for ionization fraction, level population, and line optical depths, self-consistently from the input physical parameters (HI number density $n_H$, Temperature $T$, hydrogen ionization rate $\gamma_{HI}$ and velocity gradient d$v$/d$l$), overcoming the Case B limitations. We use specific models with parameters that describe expected conditions on winds and accretion flows on CTTSs, already employed on previous works \citep[e.g,]{2013ApJ...778..148E, 2017A&A...599A.105A} as a base for comparison with our retrieved line ratios. The highest density compatible with our results vary with epoch, with $\textrm{log}(n_H)=10.2$ being compatible with the ratio of the first epoch and $\textrm{log}(n_H)=10.4$ with the second. For all other cases where $\textrm{log}(n_H) \geq 10.6$, the models predict $F_{H\alpha}/F_{H\beta} < 7$, which is below our results for either epoch. For cases where $\textrm{log}(n_H) \leq 9.2$,  $F_{H\alpha}/F_{H\beta} < 12$, which is consistent with the flux ratio during the second epoch but not with the first epoch. This suggest a HI number density on the range of $\textrm{log}(n_H)=9.4-10.2$ for HT Lup B during the first epoch, and between $\textrm{log}(n_H)=8.2-10.6$ for the second one. For the common range of $\textrm{log}(n_H)=9.4-10.2$, the line ratio of the second epoch is better reproduced by a temperature of $T=7500$ K, while $T=10000$ K is compatible with the first epoch; for any higher density case the temperatures follow an opposite trend; lower temperatures are required to reproduce the flux ratio of the first epoch, while the second epoch requires higher temperatures instead. Either the temperature change or the hydrogen density variability required to explain the difference between the first and second could be explained by the dynamic variability of the magnetosphere, as showed by simulations \cite[e.g,][]{2024MNRAS.528.2883Z} and could also be related to the stronger accretion "burst" observed during the first epoch. However, further observations are needed to properly trace the temperature and density profile of the inner disk and accretion flows, and thus these temperatures and densities should be regarded only as prelminary limits.

\subsection{Ejection rates traced by [OI] emission}
\label{subsec:ejecrate}
The interconnection between mass accretion, outflow ejection and photoevaporative winds plays an important role in the evolution of pre-main sequence stars. Accretion-driven stellar and disk winds appear to be the main source of mass loss from the disks of young stellar objects. Such mass loss from outflows and winds can effectively remove angular momentum from the disk, facilitating the infall and accretion of disk gas onto the star. For Class II sources such as HT Lup B, the presence and properties of such jets and winds are inferred mainly from the presence of strong spectroscopic tracers such as the [OI]$\lambda6300$ emission line \citep{2018A&A...620A..87M, 2023ASPC..534..567P}. Analysis of this line have shown that it can be separated into two distinct features:  high velocity components (HVC) with blueshifts from 60 to 200 km/s, originating from high velocity microjets, and low velocity components (LVC), with velocities smaller than $\sim 40$km/s, originating from low velocity disk/stellar winds \citep{2016ApJ...831..169S}. The LVC can be further divided into a Narrow Component (LVC-NC), associated with winds originating on the outer regions of the disk ($\sim 1.5$ AU), and a Broad Component (LVC-BC), tracing emission from the inner disk. Proper characterization of the [OI]$\lambda6300$ is thus a crucial step to determine the nature of the ejection processes in this type of systems.
 
To properly determine the origin of the [OI]$\lambda$6300 emission in the case of HT Lup B, we first compare the fitted line velocities for each case. Assuming that the observed velocity shifts are only a consequence of the emission mechanism itself, all datasets present a redshifted LVC, while the first epoch exhibits an additional redshifted HVC. Appearance of the HVC on the first epoch could be additional evidence for the accretion burst scenario, as this components are mostly associated with fast collimated jets possibly triggered by strong accretion events, based on the strong correlation between mass ejection and stellar mass accretion rates \citep[e.g,][]{2018A&A...609A..87N}. However, it is not possible to properly confirm this since we are not capable to spatially resolve the jet and derivation of the jet mass ejection rate is dependant on the jet extension \citep{2018A&A...609A..87N, 2018ApJ...868...28F}.

We then proceeded to analyze the LVC components for all cases, as they are indicative of a low velocity wind present at all epochs. We computed the intrinsic FWHM of the line as the square root of the quadratic difference between the measured line FWHM, calculated as FWHM$=\sigma*2\sqrt{2ln(2)}$, and the LSF FWHM, with the resulting values reported in Table \ref{tab:OIfit}. This gives a better representation of the real FWHM of the line. Adopting a criteria of FWHM $\geq 40$km/s to distinguish between a narrow and broad component \citep{2016ApJ...831..169S, 2018ApJ...868...28F}, the results for the line peak velocity and intrinsic width are consistent with a broad low velocity component (LVC-BC), which is normally associated with wind emission from the inner (within 0.5 AU) disk of the system. 

We then proceed to derive the wind mass-loss rate, $\dot{M}_{\scriptsize\textrm{wind}}$, following the prescription from \cite{2018ApJ...868...28F}:
\begin{equation}
    \dot{M}_{\textrm{wind}} = \textrm{C(T)}\left(\frac{\textrm{V}_{\textrm{wind}}}{10 \textrm{km/s}} \right) \left(\frac{l_{\textrm{wind}}}{1\textrm{AU}}\right)^{-1}
    \left(\frac{\textrm{L}_{6300}}{\textrm{L}_\odot} \right)
\end{equation}

Where V\textsubscript{wind} corresponds to the wind velocity, calculated as the de-projected peak velocity of the line for each dataset assuming an disk inclination of $44^{\circ}.9\pm4.9$, as derived from \cite{2018ApJ...869L..44K} for HT Lup B. Here l\textsubscript{wind} is the wind vertical extent, which we assume to be equal to the disk radius from which the emission is coming; this radius is calculated assuming that the broadening of the [OI] line comes solely from Keplerian rotation, and thus, the radius corresponds to the Keplerian radius from the deprojected line Half Width at Half Maximum (HWHM). The term $\textrm{L}_{6300}$ is the [OI]$\lambda6300$ line luminosity calculated from the line flux derived in Sect. \ref{subsec:OIline}. For the first epoch, we only consider the flux contained on the LVC-BC when computing the luminosity, which corresponds to $\sim 77\%$ of the total flux. Finally, C(T) is a factor dependent on the gas temperature, which we are not able to determine from our observations alone. Thus we study the limiting cases at 5000K and 10,000K, for which C(T) takes values of $2.4 \times 10^{-4}$ to $2.6 \times 10^{-5}$, respectively \citep{2018ApJ...868...28F}.

Using this approach, we derive a mean ejection rate of $8.6 \times 10^{-8} M_\odot \, yr^{-1}$, $7.5 \times 10^{-7} M_\odot \, yr^{-1}$  and $3.7 \times 10^{-7} M_\odot \, yr^{-1}$ for the first, second and third datasets, respectively, and for a temperature of 5000 K. For the 10,000 K case, the ejection rates are approximately one order of magnitude fainter at $9.3 \times 10^{-9} M_\odot \, yr^{-1}$, $8.1 \times 10^{-8} M_\odot \, yr^{-1}$  and $4.0 \times 10^{-8} M_\odot \, yr^{-1}$. All the results for the lower temperature case present higher wind ejection rate values than those expected for Class II disks, while the 10,000 K cases are more in line with observed trends \citep{2023ASPC..534..567P}. However, for all cases there are important discrepancies when compared with their accretion rates. For Class II sources, the ejection to accretion ratio normally follows $\dot{M}_{\textrm{wind}}/\dot{M}_{\textrm{acc}} \sim 0.1-1$ \citep{2014A&A...569A...5N, 2018ApJ...868...28F}, with the specific value being dependent on the temperature and wind extension of the source. Based on the the accretion rates derived on Sect.\ref{subsec:physprop} from the line luminosity, the wind ejection rates derived for both temperature cases are unreasonably large, which would suggest even higher temperatures for the wind. However, in that case the emission lines of ionized oxygen, such as [OII]$\lambda$7330, are also expected to arise and should be detectable \citep{2016ApJ...831..169S}, which is not the case. As they also do not appear on the HT Lup A spectra in Appendix \ref{App:E}, we do not expect the lack of detection of [OII]$\lambda$7330 being an effect of the HRSDI processing. Another possible source for this discrepancy are the assumed values regarding the wind geometry. We assumed the line velocity shift to be purely a tracer of the wind velocity, ignoring other possible components such as the orbital motion of HT Lup B, which could result in effectively lower or even blueshifted line velocities when corrected. Also, the keplerian radius obtained from the line HWHM are on the order of 0.02 au for the three datasets, which might be to small to be considered a realistic wind height; for example, photoevaporative winds driven by X-rays can extend at $\sim35$ AU above the disk \citep{2016MNRAS.460.3472E}, a distance over $10^3$ times higher than our assumed values. We note that considering a height increase of a factor of 10 for the first epoch results and 100 for the second epoch, over our assumed heights results on expected values of ejection rates ($\dot{M}_{\textrm{wind}}/\dot{M}_{\textrm{acc}} \sim 0.5-1$ for all epochs) for a 10,000 K temperature. Based on this we regard our results only as upper limits for the wind ejection rate of HT Lup B.

Providing an estimation of the ejection rates associated with a possible jet originating the HVC during the first epoch is beyond the scope of this work, as it requires determination of the spatial extent of the jet \citep[e.g,][]{2023A&A...670A.126F}, which we are not capable to resolve from our data. Dedicated spectroscopic follow-up at higher spectral and spatial resolution is necessary to properly determine the wind geometry and outflow behaviour of the system.

\section{Summary and Conclusions}
In this work we present the first  study of accretion/ejection processes on the close binary companion HT Lup B, based on the detection of hydrogen and [OI] emission lines. We characterize the accretion properties of the object and compare with previous estimations and with other, more classical, T Tauri objects. Our main results are summarized as follows.

\begin{itemize}
    \item We recover clearly distinct H$\alpha$ line profiles between both epochs of observation. The first epoch showcases stronger flux levels by a factor of $\sim 3$ compared to the second, and wider profiles by $\sim 60$ km/s in average. In both epochs, the line also presents an evident redshift of $70$ km/s in average.
    
    \item The retrieved linewidths for both epochs strongly point for both HT Lup A and B to be closer to a Classic T Tauri Star (CTTS) rather than a Weak-line T Tauri object (WTTS). We propose to classify both sources as CTTS, that the H$\alpha$ line properties observed for HT Lup B originate as a product of its accretion regime, and that the observed line variability is produced due to an accretion burst event during the first epoch. 
    
    \item The H$\alpha$ line profiles for both HT Lup A and B are consistent with emission due to independent accretion activity. Conversion of the integrated H$\alpha$ line fluxes into accretion rates using empirical relations for accreting T Tauri stars, indicate that HT Lup B has a level of accretion on the range of $10^{-9}-10^{-10} M_\odot \, yr^{-1}$, while HT lup A presents stronger accretion on the order of a few $10^{-9} M_\odot \, yr^{-1}$. These values are in good agreement with accretion rates derived for other accreting systems in the Lupus star forming region.
    
    \item We propose that observations of the first epoch captured evidence of an accretion burst event on HT Lup B, while observations of the second epoch correspond to a quiescent accretion phase, based on the observed line and accretion variability, which can not be completely explained with more traditional variability scenarios such as rotational modulation.
    
    \item We were also capable to retrieve H$\beta$ line emission in some of the cubes. The behaviour and variability of the line is consistent to what is seen for H$\alpha$, further supporting an H$\beta$ origin from accretion processes. Study of the flux ratio from both hydrogen lines point to an HI number density on the range of $\textrm{log}(n_H)=9.4-10.2$ common for both epochs, although the limits vary for the second epoch ($\textrm{log}(n_H)=8.2-10.6$), on the accreting flow regions of the system, with minimum temperatures between $7500-10000$ K.

    \item We retrieve forbidden [OI]$\lambda6300$ emission for most cubes at both epochs, tracing ejection processes ocurring in HT Lup B. The presence of a high velocity component on the line profile of the first epoch, which is absent from the second epoch, further support the hypothesis of an accretion burst event, as it possibly traces emission from an unresolved jet triggered by a strong accretion episode. The derived ejection rates are too large for the expected behaviour of Class II sources, mostly due to the uncertainties regarding the wind geometry.

\end{itemize}

These results are a strong argument on the instrumental capabilities of MUSE to detect and study accretion signatures on close-in companions, as well as a promising starting point on the characterization on the rather poorly studied HT Lup B object. Additional observations aiming to study the current accreting state of the system, together with its variability on timescales not explored here, are essential for a robust depiction of its accretion history. Nonetheless, HT Lup is in itself an extremely interesting case study for the evolution of more complex pre-main sequence stellar systems.

\begin{acknowledgements}
S.J.\ acknowledges support from the National Agency for Research and Development (ANID), Scholarship Program, Doctorado Becas Nacionales/2020 - 21212356. 
L.P. gratefully acknowledges support by the ANID BASAL project FB210003 and ANID FONDECYT Regular \#1221442.
S.J., L.P., G.C., M.B., acknowledge support from Programa de Cooperación Científica ECOS-ANID ECOS200049.
This research has made use of the SIMBAD database, operated at CDS, Strasbourg, France, and of NASA's Astrophysics Data System Bibliographic Services. This work benefited from the support of the project FRAME ANR-20-CE31-0012 of the French National Research Agency (ANR). This project has received funding from the European Research Council (ERC) under the European Union Horizon Europe programme (grant agreement No. 101042275, project Stellar-MADE).
\end{acknowledgements}

\bibliography{imaging}{}

\begin{thebibliography}{}
\expandafter\ifx\csname natexlab\endcsname\relax\def\natexlab#1{#1}\fi
\providecommand{\url}[1]{\href{#1}{#1}}
\providecommand{\dodoi}[1]{doi:~\href{http://doi.org/#1}{\nolinkurl{#1}}}
\providecommand{\doeprint}[1]{\href{http://ascl.net/#1}{\nolinkurl{http://ascl.net/#1}}}
\providecommand{\doarXiv}[1]{\href{https://arxiv.org/abs/#1}{\nolinkurl{https://arxiv.org/abs/#1}}}

\bibitem[{{Alcal{\'a}} {et~al.}(2014){Alcal{\'a}}, {Natta}, {Manara}, {Spezzi}, {Stelzer}, {Frasca}, {Biazzo}, {Covino}, {Randich}, {Rigliaco}, {Testi}, {Comer{\'o}n}, {Cupani}, \& {D'Elia}}]{2014A&A...561A...2A}
{Alcal{\'a}}, J.~M., {Natta}, A., {Manara}, C.~F., {et~al.} 2014, \aap, 561, A2, \dodoi{10.1051/0004-6361/201322254}

\bibitem[{{Alcal{\'a}} {et~al.}(2017){Alcal{\'a}}, {Manara}, {Natta}, {Frasca}, {Testi}, {Nisini}, {Stelzer}, {Williams}, {Antoniucci}, {Biazzo}, {Covino}, {Esposito}, {Getman}, \& {Rigliaco}}]{2017A&A...600A..20A}
{Alcal{\'a}}, J.~M., {Manara}, C.~F., {Natta}, A., {et~al.} 2017, \aap, 600, A20, \dodoi{10.1051/0004-6361/201629929}

\bibitem[{{Alencar} \& {Basri}(2000)}]{2000AJ....119.1881A}
{Alencar}, S. H.~P., \& {Basri}, G. 2000, \aj, 119, 1881, \dodoi{10.1086/301300}

\bibitem[{{Alencar} {et~al.}(2012){Alencar}, {Bouvier}, {Walter}, {Dougados}, {Donati}, {Kurosawa}, {Romanova}, {Bonfils}, {Lima}, {Massaro}, {Ibrahimov}, \& {Poretti}}]{2012A&A...541A.116A}
{Alencar}, S.~H.~P., {Bouvier}, J., {Walter}, F.~M., {et~al.} 2012, \aap, 541, A116, \dodoi{10.1051/0004-6361/201118395}

\bibitem[{{Andrews} {et~al.}(2018){Andrews}, {Huang}, {P{\'e}rez}, {Isella}, {Dullemond}, {Kurtovic}, {Guzm{\'a}n}, {Carpenter}, {Wilner}, {Zhang}, {Zhu}, {Birnstiel}, {Bai}, {Benisty}, {Hughes}, {{\"O}berg}, \& {Ricci}}]{2018ApJ...869L..41A}
{Andrews}, S.~M., {Huang}, J., {P{\'e}rez}, L.~M., {et~al.} 2018, \apjl, 869, L41, \dodoi{10.3847/2041-8213/aaf741}

\bibitem[{{Anthonioz} {et~al.}(2015){Anthonioz}, {M{\'e}nard}, {Pinte}, {Le Bouquin}, {Benisty}, {Thi}, {Absil}, {Duch{\^e}ne}, {Augereau}, {Berger}, {Casassus}, {Duvert}, {Lazareff}, {Malbet}, {Millan-Gabet}, {Schreiber}, {Traub}, \& {Zins}}]{2015A&A...574A..41A}
{Anthonioz}, F., {M{\'e}nard}, F., {Pinte}, C., {et~al.} 2015, \aap, 574, A41, \dodoi{10.1051/0004-6361/201424520}

\bibitem[{{Antoniucci} {et~al.}(2014){Antoniucci}, {Garc{\'\i}a L{\'o}pez}, {Nisini}, {Caratti o Garatti}, {Giannini}, \& {Lorenzetti}}]{2014A&A...572A..62A}
{Antoniucci}, S., {Garc{\'\i}a L{\'o}pez}, R., {Nisini}, B., {et~al.} 2014, \aap, 572, A62, \dodoi{10.1051/0004-6361/201423929}

\bibitem[{{Antoniucci} {et~al.}(2017){Antoniucci}, {Nisini}, {Giannini}, {Rigliaco}, {Alcal{\'a}}, {Natta}, \& {Stelzer}}]{2017A&A...599A.105A}
{Antoniucci}, S., {Nisini}, B., {Giannini}, T., {et~al.} 2017, \aap, 599, A105, \dodoi{10.1051/0004-6361/201629683}

\bibitem[{{Arsenault} {et~al.}(2008){Arsenault}, {Madec}, {Hubin}, {Paufique}, {Stroebele}, {Soenke}, {Donaldson}, {Fedrigo}, {Oberti}, {Tordo}, {Downing}, {Kiekebusch}, {Conzelmann}, {Duchateau}, {Jost}, {Hackenberg}, {Bonaccini Calia}, {Delabre}, {Stuik}, {Biasi}, {Gallieni}, {Lazzarini}, {Lelouarn}, \& {Glindeman}}]{2008SPIE.7015E..24A}
{Arsenault}, R., {Madec}, P.~Y., {Hubin}, N., {et~al.} 2008, in Society of Photo-Optical Instrumentation Engineers (SPIE) Conference Series, Vol. 7015, Adaptive Optics Systems, ed. N.~{Hubin}, C.~E. {Max}, \& P.~L. {Wizinowich}, 701524, \dodoi{10.1117/12.790359}

\bibitem[{{Artymowicz} \& {Lubow}(1994)}]{1994ApJ...421..651A}
{Artymowicz}, P., \& {Lubow}, S.~H. 1994, \apj, 421, 651, \dodoi{10.1086/173679}

\bibitem[{{Bacon} {et~al.}(2010){Bacon}, {Accardo}, {Adjali}, {Anwand}, {Bauer}, {Biswas}, {Blaizot}, {Boudon}, {Brau-Nogue}, {Brinchmann}, {Caillier}, {Capoani}, {Carollo}, {Contini}, {Couderc}, {Daguis{\'e}}, {Deiries}, {Delabre}, {Dreizler}, {Dubois}, {Dupieux}, {Dupuy}, {Emsellem}, {Fechner}, {Fleischmann}, {Fran{\c{c}}ois}, {Gallou}, {Gharsa}, {Glindemann}, {Gojak}, {Guiderdoni}, {Hansali}, {Hahn}, {Jarno}, {Kelz}, {Koehler}, {Kosmalski}, {Laurent}, {Le Floch}, {Lilly}, {Lizon}, {Loupias}, {Manescau}, {Monstein}, {Nicklas}, {Olaya}, {Pares}, {Pasquini}, {P{\'e}contal-Rousset}, {Pell{\'o}}, {Petit}, {Popow}, {Reiss}, {Remillieux}, {Renault}, {Roth}, {Rupprecht}, {Serre}, {Schaye}, {Soucail}, {Steinmetz}, {Streicher}, {Stuik}, {Valentin}, {Vernet}, {Weilbacher}, {Wisotzki}, \& {Yerle}}]{2010SPIE.7735E..08B}
{Bacon}, R., {Accardo}, M., {Adjali}, L., {et~al.} 2010, in Society of Photo-Optical Instrumentation Engineers (SPIE) Conference Series, Vol. 7735, Ground-based and Airborne Instrumentation for Astronomy III, ed. I.~S. {McLean}, S.~K. {Ramsay}, \& H.~{Takami}, 773508, \dodoi{10.1117/12.856027}

\bibitem[{{Baraffe} {et~al.}(2015){Baraffe}, {Homeier}, {Allard}, \& {Chabrier}}]{2015A&A...577A..42B}
{Baraffe}, I., {Homeier}, D., {Allard}, F., \& {Chabrier}, G. 2015, \aap, 577, A42, \dodoi{10.1051/0004-6361/201425481}

\bibitem[{{Bary} \& {Petersen}(2014)}]{2014ApJ...792...64B}
{Bary}, J.~S., \& {Petersen}, M.~S. 2014, \apj, 792, 64, \dodoi{10.1088/0004-637X/792/1/64}

\bibitem[{{Bate} \& {Bonnell}(1997)}]{1997MNRAS.285...33B}
{Bate}, M.~R., \& {Bonnell}, I.~A. 1997, \mnras, 285, 33, \dodoi{10.1093/mnras/285.1.33}

\bibitem[{{Blunt} {et~al.}(2017){Blunt}, {Nielsen}, {De Rosa}, {Konopacky}, {Ryan}, {Wang}, {Pueyo}, {Rameau}, {Marois}, {Marchis}, {Macintosh}, {Graham}, {Duch{\^e}ne}, \& {Schneider}}]{2017AJ....153..229B}
{Blunt}, S., {Nielsen}, E.~L., {De Rosa}, R.~J., {et~al.} 2017, \aj, 153, 229, \dodoi{10.3847/1538-3881/aa6930}

\bibitem[{{Calvet} {et~al.}(2000){Calvet}, {Hartmann}, \& {Strom}}]{2000prpl.conf..377C}
{Calvet}, N., {Hartmann}, L., \& {Strom}, S.~E. 2000, in Protostars and Planets IV, ed. V.~{Mannings}, A.~P. {Boss}, \& S.~S. {Russell}, 377, \dodoi{10.48550/arXiv.astro-ph/9902335}

\bibitem[{{Ceppi} {et~al.}(2022){Ceppi}, {Cuello}, {Lodato}, {Clarke}, {Toci}, \& {Price}}]{2022MNRAS.514..906C}
{Ceppi}, S., {Cuello}, N., {Lodato}, G., {et~al.} 2022, \mnras, 514, 906, \dodoi{10.1093/mnras/stac1390}

\bibitem[{{Cieza} {et~al.}(2013){Cieza}, {Olofsson}, {Harvey}, {Evans}, {Najita}, {Henning}, {Mer{\'\i}n}, {Liebhart}, {G{\"u}del}, {Augereau}, \& {Pinte}}]{2013ApJ...762..100C}
{Cieza}, L.~A., {Olofsson}, J., {Harvey}, P.~M., {et~al.} 2013, \apj, 762, 100, \dodoi{10.1088/0004-637X/762/2/100}

\bibitem[{{Close}(2020)}]{2020AJ....160..221C}
{Close}, L.~M. 2020, \aj, 160, 221, \dodoi{10.3847/1538-3881/abb375}

\bibitem[{{Costigan} {et~al.}(2014){Costigan}, {Vink}, {Scholz}, {Ray}, \& {Testi}}]{2014MNRAS.440.3444C}
{Costigan}, G., {Vink}, J.~S., {Scholz}, A., {Ray}, T., \& {Testi}, L. 2014, \mnras, 440, 3444, \dodoi{10.1093/mnras/stu529}

\bibitem[{{Daemgen} {et~al.}(2012){Daemgen}, {Correia}, \& {Petr-Gotzens}}]{2012A&A...540A..46D}
{Daemgen}, S., {Correia}, S., \& {Petr-Gotzens}, M.~G. 2012, \aap, 540, A46, \dodoi{10.1051/0004-6361/201118314}

\bibitem[{{Duch{\^e}ne} {et~al.}(2024){Duch{\^e}ne}, {LeBouquin}, {M{\'e}nard}, {Cuello}, {Toci}, \& {Langlois}}]{2024arXiv240402469D}
{Duch{\^e}ne}, G., {LeBouquin}, J.-B., {M{\'e}nard}, F., {et~al.} 2024, arXiv e-prints, arXiv:2404.02469, \dodoi{10.48550/arXiv.2404.02469}

\bibitem[{{Dvorak}(1982)}]{1982OAWMN.191..423D}
{Dvorak}, R. 1982, Oesterreichische Akademie Wissenschaften Mathematisch naturwissenschaftliche Klasse Sitzungsberichte Abteilung, 191, 423

\bibitem[{{Edwards} {et~al.}(2013){Edwards}, {Kwan}, {Fischer}, {Hillenbrand}, {Finn}, {Fedorenko}, \& {Feng}}]{2013ApJ...778..148E}
{Edwards}, S., {Kwan}, J., {Fischer}, W., {et~al.} 2013, \apj, 778, 148, \dodoi{10.1088/0004-637X/778/2/148}

\bibitem[{{Ercolano} \& {Owen}(2016)}]{2016MNRAS.460.3472E}
{Ercolano}, B., \& {Owen}, J.~E. 2016, \mnras, 460, 3472, \dodoi{10.1093/mnras/stw1179}

\bibitem[{{Fang} {et~al.}(2018){Fang}, {Pascucci}, {Edwards}, {Gorti}, {Banzatti}, {Flock}, {Hartigan}, {Herczeg}, \& {Dupree}}]{2018ApJ...868...28F}
{Fang}, M., {Pascucci}, I., {Edwards}, S., {et~al.} 2018, \apj, 868, 28, \dodoi{10.3847/1538-4357/aae780}

\bibitem[{{Fiorellino} {et~al.}(2022){Fiorellino}, {Park}, {K{\'o}sp{\'a}l}, \& {{\'A}brah{\'a}m}}]{2022ApJ...928...81F}
{Fiorellino}, E., {Park}, S., {K{\'o}sp{\'a}l}, {\'A}., \& {{\'A}brah{\'a}m}, P. 2022, \apj, 928, 81, \dodoi{10.3847/1538-4357/ac4790}

\bibitem[{{Fitzpatrick} \& {Massa}(2007)}]{2007ApJ...663..320F}
{Fitzpatrick}, E.~L., \& {Massa}, D. 2007, \apj, 663, 320, \dodoi{10.1086/518158}

\bibitem[{{Flores-Rivera} {et~al.}(2023){Flores-Rivera}, {Flock}, {Kurtovic}, {Husemann}, {Banzatti}, {Ringqvist}, {Kamann}, {M{\"u}ller}, {Fendt}, {Garc{\'\i}a Lopez}, {Marleau}, {Henning}, {Carrasco-Gonz{\'a}lez}, {van Boekel}, {Keppler}, {Launhardt}, \& {Aoyama}}]{2023A&A...670A.126F}
{Flores-Rivera}, L., {Flock}, M., {Kurtovic}, N.~T., {et~al.} 2023, \aap, 670, A126, \dodoi{10.1051/0004-6361/202141664}

\bibitem[{{Foreman-Mackey} {et~al.}(2013){Foreman-Mackey}, {Hogg}, {Lang}, \& {Goodman}}]{2013PASP..125..306F}
{Foreman-Mackey}, D., {Hogg}, D.~W., {Lang}, D., \& {Goodman}, J. 2013, \pasp, 125, 306, \dodoi{10.1086/670067}

\bibitem[{{Frasca} {et~al.}(2015){Frasca}, {Biazzo}, {Lanzafame}, {Alcal{\'a}}, {Brugaletta}, {Klutsch}, {Stelzer}, {Sacco}, {Spina}, {Jeffries}, {Montes}, {Alfaro}, {Barentsen}, {Bonito}, {Gameiro}, {L{\'o}pez-Santiago}, {Pace}, {Pasquini}, {Prisinzano}, {Sousa}, {Gilmore}, {Randich}, {Micela}, {Bragaglia}, {Flaccomio}, {Bayo}, {Costado}, {Franciosini}, {Hill}, {Hourihane}, {Jofr{\'e}}, {Lardo}, {Maiorca}, {Masseron}, {Morbidelli}, \& {Worley}}]{2015A&A...575A...4F}
{Frasca}, A., {Biazzo}, K., {Lanzafame}, A.~C., {et~al.} 2015, \aap, 575, A4, \dodoi{10.1051/0004-6361/201424409}

\bibitem[{{Gaia Collaboration} {et~al.}(2018){Gaia Collaboration}, {Katz}, {Antoja}, {Romero-G{\'o}mez}, {Drimmel}, {Reyl{\'e}}, {Seabroke}, {Soubiran}, {Babusiaux}, {Di Matteo}, {Figueras}, {Poggio}, {Robin}, {Evans}, {Brown}, {Vallenari}, {Prusti}, {de Bruijne}, {Bailer-Jones}, {Biermann}, {Eyer}, {Jansen}, {Jordi}, {Klioner}, {Lammers}, {Lindegren}, {Luri}, {Mignard}, {Panem}, {Pourbaix}, {Randich}, {Sartoretti}, {Siddiqui}, {van Leeuwen}, {Walton}, {Arenou}, {Bastian}, {Cropper}, {Lattanzi}, {Bakker}, {Cacciari}, {Casta n}, {Chaoul}, {Cheek}, {De Angeli}, {Fabricius}, {Guerra}, {Holl}, {Masana}, {Messineo}, {Mowlavi}, {Nienartowicz}, {Panuzzo}, {Portell}, {Riello}, {Tanga}, {Th{\'e}venin}, {Gracia-Abril}, {Comoretto}, {Garcia-Reinaldos}, {Teyssier}, {Altmann}, {Andrae}, {Audard}, {Bellas-Velidis}, {Benson}, {Berthier}, {Blomme}, {Burgess}, {Busso}, {Carry}, {Cellino}, {Clementini}, {Clotet}, {Creevey}, {Davidson}, {De Ridder}, {Delchambre}, {Dell'Oro}, {Ducourant}, {Fern{\'a}ndez-Hern{\'a}ndez},
  {Fouesneau}, {Fr{\'e}mat}, {Galluccio}, {Garc{\'\i}a-Torres}, {Gonz{\'a}lez-N{\'u}{\~n}ez}, {Gonz{\'a}lez-Vidal}, {Gosset}, {Guy}, {Halbwachs}, {Hambly}, {Harrison}, {Hern{\'a}ndez}, {Hestroffer}, {Hodgkin}, {Hutton}, {Jasniewicz}, {Jean-Antoine-Piccolo}, {Jordan}, {Korn}, {Krone-Martins}, {Lanzafame}, {Lebzelter}, {L{\"o}ffler}, {Manteiga}, {Marrese}, {Mart{\'\i}n-Fleitas}, {Moitinho}, {Mora}, {Muinonen}, {Osinde}, {Pancino}, {Pauwels}, {Petit}, {Recio-Blanco}, {Richards}, {Rimoldini}, {Sarro}, {Siopis}, {Smith}, {Sozzetti}, {S{\"u}veges}, {Torra}, {van Reeven}, {Abbas}, {Abreu Aramburu}, {Accart}, {Aerts}, {Altavilla}, {{\'A}lvarez}, {Alvarez}, {Alves}, {Anderson}, {Andrei}, {Anglada Varela}, {Antiche}, {Arcay}, {Astraatmadja}, {Bach}, {Baker}, {Balaguer-N{\'u}{\~n}ez}, {Balm}, {Barache}, {Barata}, {Barbato}, {Barblan}, {Barklem}, {Barrado}, {Barros}, {Barstow}, {Bartholom{\'e} Mu{\~n}oz}, {Bassilana}, {Becciani}, {Bellazzini}, {Berihuete}, {Bertone}, {Bianchi}, {Bienaym{\'e}}, {Blanco-Cuaresma}, {Boch},
  {Boeche}, {Bombrun}, {Borrachero}, {Bossini}, {Bouquillon}, {Bourda}, {Bragaglia}, {Bramante}, {Breddels}, {Bressan}, {Brouillet}, {Br{\"u}semeister}, {Brugaletta}, {Bucciarelli}, {Burlacu}, {Busonero}, {Butkevich}, {Buzzi}, {Caffau}, {Cancelliere}, {Cannizzaro}, {Cantat-Gaudin}, {Carballo}, {Carlucci}, {Carrasco}, {Casamiquela}, {Castellani}, {Castro-Ginard}, {Charlot}, {Chemin}, {Chiavassa}, {Cocozza}, {Costigan}, {Cowell}, {Crifo}, {Crosta}, {Crowley}, {Cuypers}, {Dafonte}, {Damerdji}, {Dapergolas}, {David}, {David}, {de Laverny}, {De Luise}, {De March}, {de Souza}, {de Torres}, {Debosscher}, {del Pozo}, {Delbo}, {Delgado}, {Delgado}, {Diakite}, {Diener}, {Distefano}, {Dolding}, {Drazinos}, {Dur{\'a}n}, {Edvardsson}, {Enke}, {Eriksson}, {Esquej}, {Eynard Bontemps}, {Fabre}, {Fabrizio}, {Faigler}, {Falc a}, {Farr{\`a}s Casas}, {Federici}, {Fedorets}, {Fernique}, {Filippi}, {Findeisen}, {Fonti}, {Fraile}, {Fraser}, {Fr{\'e}zouls}, {Gai}, {Galleti}, {Garabato}, {Garc{\'\i}a-Sedano}, {Garofalo}, {Garralda},
  {Gavel}, {Gavras}, {Gerssen}, {Geyer}, {Giacobbe}, {Gilmore}, {Girona}, {Giuffrida}, {Glass}, {Gomes}, {Granvik}, {Gueguen}, {Guerrier}, {Guiraud}, {Guti{\'e}}, {Haigron}, {Hatzidimitriou}, {Hauser}, {Haywood}, {Heiter}, {Helmi}, {Heu}, {Hilger}, {Hobbs}, {Hofmann}, {Holland}, {Huckle}, {Hypki}, {Icardi}, {Jan{\ss}en}, {Jevardat de Fombelle}, {Jonker}, {Juh{\'a}sz}, {Julbe}, {Karampelas}, {Kewley}, {Klar}, {Kochoska}, {Kohley}, {Kolenberg}, {Kontizas}, {Kontizas}, {Koposov}, {Kordopatis}, {Kostrzewa-Rutkowska}, {Koubsky}, {Lambert}, {Lanza}, {Lasne}, {Lavigne}, {Le Fustec}, {Le Poncin-Lafitte}, {Lebreton}, {Leccia}, {Leclerc}, {Lecoeur-Taibi}, {Lenhardt}, {Leroux}, {Liao}, {Licata}, {Lindstr{\o}m}, {Lister}, {Livanou}, {Lobel}, {L{\'o}pez}, {Managau}, {Mann}, {Mantelet}, {Marchal}, {Marchant}, {Marconi}, {Marinoni}, {Marschalk{\'o}}, {Marshall}, {Martino}, {Marton}, {Mary}, {Massari}, {Matijevi{\v{c}}}, {Mazeh}, {McMillan}, {Messina}, {Michalik}, {Millar}, {Molina}, {Molinaro}, {Moln{\'a}r}, {Montegriffo},
  {Mor}, {Morbidelli}, {Morel}, {Morris}, {Mulone}, {Muraveva}, {Musella}, {Nelemans}, {Nicastro}, {Noval}, {O'Mullane}, {Ord{\'e}novic}, {Ord{\'o}{\~n}ez-Blanco}, {Osborne}, {Pagani}, {Pagano}, {Pailler}, {Palacin}, {Palaversa}, {Panahi}, {Pawlak}, {Piersimoni}, {Pineau}, {Plachy}, {Plum}, {Poujoulet}, {Pr{\v{s}}a}, {Pulone}, {Racero}, {Ragaini}, {Rambaux}, {Ramos-Lerate}, {Regibo}, {Riclet}, {Ripepi}, {Riva}, {Rivard}, {Rixon}, {Roegiers}, {Roelens}, {Rowell}, {Royer}, {Ruiz-Dern}, {Sadowski}, {Sagrist{\`a} Sell{\'e}s}, {Sahlmann}, {Salgado}, {Salguero}, {Sanna}, {Santana-Ros}, {Sarasso}, {Savietto}, {Schultheis}, {Sciacca}, {Segol}, {Segovia}, {S{\'e}gransan}, {Shih}, {Siltala}, {Silva}, {Smart}, {Smith}, {Solano}, {Solitro}, {Sordo}, {Soria Nieto}, {Souchay}, {Spagna}, {Spoto}, {Stampa}, {Steele}, {Steidelm{\"u}ller}, {Stephenson}, {Stoev}, {Suess}, {Surdej}, {Szabados}, {Szegedi-Elek}, {Tapiador}, {Taris}, {Tauran}, {Taylor}, {Teixeira}, {Terrett}, {Teyssandier}, {Thuillot}, {Titarenko}, {Torra Clotet},
  {Turon}, {Ulla}, {Utrilla}, {Uzzi}, {Vaillant}, {Valentini}, {Valette}, {van Elteren}, {Van Hemelryck}, {van Leeuwen}, {Vaschetto}, {Vecchiato}, {Veljanoski}, {Viala}, {Vicente}, {Vogt}, {von Essen}, {Voss}, {Votruba}, {Voutsinas}, {Walmsley}, {Weiler}, {Wertz}, {Wevers}, {Wyrzykowski}, {Yoldas}, {{\v{Z}}erjal}, {Ziaeepour}, {Zorec}, {Zschocke}, {Zucker}, {Zurbach}, \& {Zwitter}}]{2018A&A...616A..11G}
{Gaia Collaboration}, {Katz}, D., {Antoja}, T., {et~al.} 2018, \aap, 616, A11, \dodoi{10.1051/0004-6361/201832865}

\bibitem[{{Ghez} {et~al.}(1997){Ghez}, {White}, \& {Simon}}]{1997ApJ...490..353G}
{Ghez}, A.~M., {White}, R.~J., \& {Simon}, M. 1997, \apj, 490, 353, \dodoi{10.1086/304856}

\bibitem[{{Gras-Vel{\'a}zquez} \& {Ray}(2005)}]{2005A&A...443..541G}
{Gras-Vel{\'a}zquez}, {\`A}., \& {Ray}, T.~P. 2005, \aap, 443, 541, \dodoi{10.1051/0004-6361:20042397}

\bibitem[{{Gullbring}(1994)}]{1994A&A...287..131G}
{Gullbring}, E. 1994, \aap, 287, 131

\bibitem[{{Haffert} {et~al.}(2019){Haffert}, {Bohn}, {de Boer}, {Snellen}, {Brinchmann}, {Girard}, {Keller}, \& {Bacon}}]{2019NatAs...3..749H}
{Haffert}, S.~Y., {Bohn}, A.~J., {de Boer}, J., {et~al.} 2019, Nature Astronomy, 3, 749, \dodoi{10.1038/s41550-019-0780-5}

\bibitem[{{Herczeg} \& {Hillenbrand}(2008)}]{2008ApJ...681..594H}
{Herczeg}, G.~J., \& {Hillenbrand}, L.~A. 2008, \apj, 681, 594, \dodoi{10.1086/586728}

\bibitem[{{Hirsh} {et~al.}(2020){Hirsh}, {Price}, {Gonzalez}, {Ubeira-Gabellini}, \& {Ragusa}}]{2020MNRAS.498.2936H}
{Hirsh}, K., {Price}, D.~J., {Gonzalez}, J.-F., {Ubeira-Gabellini}, M.~G., \& {Ragusa}, E. 2020, \mnras, 498, 2936, \dodoi{10.1093/mnras/staa2536}

\bibitem[{{Hughes} {et~al.}(1994){Hughes}, {Hartigan}, {Krautter}, \& {Kelemen}}]{1994AJ....108.1071H}
{Hughes}, J., {Hartigan}, P., {Krautter}, J., \& {Kelemen}, J. 1994, \aj, 108, 1071, \dodoi{10.1086/117135}

\bibitem[{{Jones, A.} {et~al.}(2013){Jones, A.}, {Noll, S.}, {Kausch, W.}, {Szyszka, C.}, \& {Kimeswenger, S.}}]{refId1}
{Jones, A.}, {Noll, S.}, {Kausch, W.}, {Szyszka, C.}, \& {Kimeswenger, S.} 2013, A\&A, 560, A91, \dodoi{10.1051/0004-6361/201322433}

\bibitem[{{Kraus} {et~al.}(2012){Kraus}, {Ireland}, {Hillenbrand}, \& {Martinache}}]{2012ApJ...745...19K}
{Kraus}, A.~L., {Ireland}, M.~J., {Hillenbrand}, L.~A., \& {Martinache}, F. 2012, \apj, 745, 19, \dodoi{10.1088/0004-637X/745/1/19}

\bibitem[{{Kurosawa} {et~al.}(2006){Kurosawa}, {Harries}, \& {Symington}}]{2006MNRAS.370..580K}
{Kurosawa}, R., {Harries}, T.~J., \& {Symington}, N.~H. 2006, \mnras, 370, 580, \dodoi{10.1111/j.1365-2966.2006.10527.x}

\bibitem[{{Kurosawa} \& {Romanova}(2012)}]{2012MNRAS.426.2901K}
{Kurosawa}, R., \& {Romanova}, M.~M. 2012, \mnras, 426, 2901, \dodoi{10.1111/j.1365-2966.2012.21853.x}

\bibitem[{{Kurosawa} {et~al.}(2011){Kurosawa}, {Romanova}, \& {Harries}}]{2011MNRAS.416.2623K}
{Kurosawa}, R., {Romanova}, M.~M., \& {Harries}, T.~J. 2011, \mnras, 416, 2623, \dodoi{10.1111/j.1365-2966.2011.19216.x}

\bibitem[{Kurtovic(2019)}]{Kurtovic_2019}
Kurtovic, N. 2019, Characterizing-substructures-and-interactions-in-the-disks ...
\newblock \url{https://repositorio.uchile.cl/bitstream/handle/2250/173021/Characterizing-substructures-and-interactions-in-the-disks-around-multiple.pdf?sequence=1&isAllowed=y}

\bibitem[{{Kurtovic} {et~al.}(2018){Kurtovic}, {P{\'e}rez}, {Benisty}, {Zhu}, {Zhang}, {Huang}, {Andrews}, {Dullemond}, {Isella}, {Bai}, {Carpenter}, {Guzm{\'a}n}, {Ricci}, \& {Wilner}}]{2018ApJ...869L..44K}
{Kurtovic}, N.~T., {P{\'e}rez}, L.~M., {Benisty}, M., {et~al.} 2018, \apjl, 869, L44, \dodoi{10.3847/2041-8213/aaf746}

\bibitem[{{Kwan} \& {Fischer}(2011)}]{2011MNRAS.411.2383K}
{Kwan}, J., \& {Fischer}, W. 2011, \mnras, 411, 2383, \dodoi{10.1111/j.1365-2966.2010.17863.x}

\bibitem[{{Lai} \& {Mu{\~n}oz}(2022)}]{2022arXiv221100028L}
{Lai}, D., \& {Mu{\~n}oz}, D.~J. 2022, arXiv e-prints, arXiv:2211.00028, \dodoi{10.48550/arXiv.2211.00028}

\bibitem[{{Lai} \& {Mu{\~n}oz}(2023)}]{2023ARA&A..61..517L}
---. 2023, \araa, 61, 517, \dodoi{10.1146/annurev-astro-052622-022933}

\bibitem[{{Liefke} {et~al.}(2010){Liefke}, {Fuhrmeister}, \& {Schmitt}}]{2010A&A...514A..94L}
{Liefke}, C., {Fuhrmeister}, B., \& {Schmitt}, J.~H.~M.~M. 2010, \aap, 514, A94, \dodoi{10.1051/0004-6361/201014012}

\bibitem[{{Mawet} {et~al.}(2014){Mawet}, {Milli}, {Wahhaj}, {Pelat}, {Absil}, {Delacroix}, {Boccaletti}, {Kasper}, {Kenworthy}, {Marois}, {Mennesson}, \& {Pueyo}}]{2014ApJ...792...97M}
{Mawet}, D., {Milli}, J., {Wahhaj}, Z., {et~al.} 2014, \apj, 792, 97, \dodoi{10.1088/0004-637X/792/2/97}

\bibitem[{{McGinnis} {et~al.}(2018){McGinnis}, {Dougados}, {Alencar}, {Bouvier}, \& {Cabrit}}]{2018A&A...620A..87M}
{McGinnis}, P., {Dougados}, C., {Alencar}, S.~H.~P., {Bouvier}, J., \& {Cabrit}, S. 2018, \aap, 620, A87, \dodoi{10.1051/0004-6361/201731629}

\bibitem[{{Moe} \& {Di Stefano}(2017)}]{2017ApJS..230...15M}
{Moe}, M., \& {Di Stefano}, R. 2017, \apjs, 230, 15, \dodoi{10.3847/1538-4365/aa6fb6}

\bibitem[{{Muzerolle} {et~al.}(2013){Muzerolle}, {Furlan}, {Flaherty}, {Balog}, \& {Gutermuth}}]{2013Natur.493..378M}
{Muzerolle}, J., {Furlan}, E., {Flaherty}, K., {Balog}, Z., \& {Gutermuth}, R. 2013, \nat, 493, 378, \dodoi{10.1038/nature11746}

\bibitem[{{Natta} {et~al.}(2014){Natta}, {Testi}, {Alcal{\'a}}, {Rigliaco}, {Covino}, {Stelzer}, \& {D'Elia}}]{2014A&A...569A...5N}
{Natta}, A., {Testi}, L., {Alcal{\'a}}, J.~M., {et~al.} 2014, \aap, 569, A5, \dodoi{10.1051/0004-6361/201424136}

\bibitem[{{Natta} {et~al.}(2004){Natta}, {Testi}, {Muzerolle}, {Randich}, {Comer{\'o}n}, \& {Persi}}]{2004A&A...424..603N}
{Natta}, A., {Testi}, L., {Muzerolle}, J., {et~al.} 2004, \aap, 424, 603, \dodoi{10.1051/0004-6361:20040356}

\bibitem[{{Nisini} {et~al.}(2018){Nisini}, {Antoniucci}, {Alcal{\'a}}, {Giannini}, {Manara}, {Natta}, {Fedele}, \& {Biazzo}}]{2018A&A...609A..87N}
{Nisini}, B., {Antoniucci}, S., {Alcal{\'a}}, J.~M., {et~al.} 2018, \aap, 609, A87, \dodoi{10.1051/0004-6361/201730834}

\bibitem[{{Noll, S.} {et~al.}(2012){Noll, S.}, {Kausch, W.}, {Barden, M.}, {Jones, A. M.}, {Szyszka, C.}, {Kimeswenger, S.}, \& {Vinther, J.}}]{refId0}
{Noll, S.}, {Kausch, W.}, {Barden, M.}, {et~al.} 2012, A\&A, 543, A92, \dodoi{10.1051/0004-6361/201219040}

\bibitem[{{Offner} {et~al.}(2022){Offner}, {Moe}, {Kratter}, {Sadavoy}, {Jensen}, \& {Tobin}}]{2022arXiv220310066O}
{Offner}, S. S.~R., {Moe}, M., {Kratter}, K.~M., {et~al.} 2022, arXiv e-prints, arXiv:2203.10066, \dodoi{10.48550/arXiv.2203.10066}

\bibitem[{{Pascucci} {et~al.}(2023){Pascucci}, {Cabrit}, {Edwards}, {Gorti}, {Gressel}, \& {Suzuki}}]{2023ASPC..534..567P}
{Pascucci}, I., {Cabrit}, S., {Edwards}, S., {et~al.} 2023, in Astronomical Society of the Pacific Conference Series, Vol. 534, Astronomical Society of the Pacific Conference Series, ed. S.~{Inutsuka}, Y.~{Aikawa}, T.~{Muto}, K.~{Tomida}, \& M.~{Tamura}, 567, \dodoi{10.48550/arXiv.2203.10068}

\bibitem[{{Petrus} {et~al.}(2021){Petrus}, {Bonnefoy}, {Chauvin}, {Charnay}, {Marleau}, {Gratton}, {Lagrange}, {Rameau}, {Mordasini}, {Nowak}, {Delorme}, {Boccaletti}, {Carlotti}, {Houll{\'e}}, {Vigan}, {Allard}, {Desidera}, {D'Orazi}, {Hoeijmakers}, {Wyttenbach}, \& {Lavie}}]{2021A&A...648A..59P}
{Petrus}, S., {Bonnefoy}, M., {Chauvin}, G., {et~al.} 2021, \aap, 648, A59, \dodoi{10.1051/0004-6361/202038914}

\bibitem[{{Raghavan} {et~al.}(2010){Raghavan}, {McAlister}, {Henry}, {Latham}, {Marcy}, {Mason}, {Gies}, {White}, \& {ten Brummelaar}}]{2010ApJS..190....1R}
{Raghavan}, D., {McAlister}, H.~A., {Henry}, T.~J., {et~al.} 2010, \apjs, 190, 1, \dodoi{10.1088/0067-0049/190/1/1}

\bibitem[{{Ragusa} {et~al.}(2020){Ragusa}, {Alexander}, {Calcino}, {Hirsh}, \& {Price}}]{2020MNRAS.499.3362R}
{Ragusa}, E., {Alexander}, R., {Calcino}, J., {Hirsh}, K., \& {Price}, D.~J. 2020, \mnras, 499, 3362, \dodoi{10.1093/mnras/staa2954}

\bibitem[{{Rich} {et~al.}(2022){Rich}, {Monnier}, {Aarnio}, {Laws}, {Setterholm}, {Wilner}, {Calvet}, {Harries}, {Miller}, {Davies}, {Adams}, {Andrews}, {Bae}, {Espaillat}, {Greenbaum}, {Hinkley}, {Kraus}, {Hartmann}, {Isella}, {McClure}, {Oppenheimer}, {P{\'e}rez}, \& {Zhu}}]{2022AJ....164..109R}
{Rich}, E.~A., {Monnier}, J.~D., {Aarnio}, A., {et~al.} 2022, \aj, 164, 109, \dodoi{10.3847/1538-3881/ac7be4}

\bibitem[{{Rigon} {et~al.}(2017){Rigon}, {Scholz}, {Anderson}, \& {West}}]{2017MNRAS.465.3889R}
{Rigon}, L., {Scholz}, A., {Anderson}, D., \& {West}, R. 2017, \mnras, 465, 3889, \dodoi{10.1093/mnras/stw2977}

\bibitem[{{Robinson} {et~al.}(1990){Robinson}, {Cram}, \& {Giampapa}}]{1990ApJS...74..891R}
{Robinson}, R.~D., {Cram}, L.~E., \& {Giampapa}, M.~S. 1990, \apjs, 74, 891, \dodoi{10.1086/191525}

\bibitem[{{Simon} {et~al.}(2016){Simon}, {Pascucci}, {Edwards}, {Feng}, {Gorti}, {Hollenbach}, {Rigliaco}, \& {Keane}}]{2016ApJ...831..169S}
{Simon}, M.~N., {Pascucci}, I., {Edwards}, S., {et~al.} 2016, \apj, 831, 169, \dodoi{10.3847/0004-637X/831/2/169}

\bibitem[{{Stauffer} {et~al.}(2014){Stauffer}, {Cody}, {Baglin}, {Alencar}, {Rebull}, {Hillenbrand}, {Venuti}, {Turner}, {Carpenter}, {Plavchan}, {Findeisen}, {Carey}, {Terebey}, {Morales-Calder{\'o}n}, {Bouvier}, {Micela}, {Flaccomio}, {Song}, {Gutermuth}, {Hartmann}, {Calvet}, {Whitney}, {Barrado}, {Vrba}, {Covey}, {Herbst}, {Furesz}, {Aigrain}, \& {Favata}}]{2014AJ....147...83S}
{Stauffer}, J., {Cody}, A.~M., {Baglin}, A., {et~al.} 2014, \aj, 147, 83, \dodoi{10.1088/0004-6256/147/4/83}

\bibitem[{{Stauffer} {et~al.}(2016){Stauffer}, {Cody}, {Rebull}, {Hillenbrand}, {Turner}, {Carpenter}, {Carey}, {Terebey}, {Morales-Calder{\'o}n}, {Alencar}, {McGinnis}, {Sousa}, {Bouvier}, {Venuti}, {Hartmann}, {Calvet}, {Micela}, {Flaccomio}, {Song}, {Gutermuth}, {Barrado}, {Vrba}, {Covey}, {Herbst}, {Gillen}, {Medeiros Guimar{\~a}es}, {Bouy}, \& {Favata}}]{2016AJ....151...60S}
{Stauffer}, J., {Cody}, A.~M., {Rebull}, L., {et~al.} 2016, \aj, 151, 60, \dodoi{10.3847/0004-6256/151/3/60}

\bibitem[{{Str{\"o}bele} {et~al.}(2012){Str{\"o}bele}, {La Penna}, {Arsenault}, {Conzelmann}, {Delabre}, {Duchateau}, {Dorn}, {Fedrigo}, {Hubin}, {Quentin}, {Jolley}, {Kiekebusch}, {Kirchbauer}, {Klein}, {Kolb}, {Kuntschner}, {Le Louarn}, {Lizon}, {Madec}, {Pettazzi}, {Soenke}, {Tordo}, {Vernet}, \& {Muradore}}]{2012SPIE.8447E..37S}
{Str{\"o}bele}, S., {La Penna}, P., {Arsenault}, R., {et~al.} 2012, in Society of Photo-Optical Instrumentation Engineers (SPIE) Conference Series, Vol. 8447, Adaptive Optics Systems III, ed. B.~L. {Ellerbroek}, E.~{Marchetti}, \& J.-P. {V{\'e}ran}, 844737, \dodoi{10.1117/12.926110}

\bibitem[{{Tofflemire} {et~al.}(2017{\natexlab{a}}){Tofflemire}, {Mathieu}, {Ardila}, {Akeson}, {Ciardi}, {Johns-Krull}, {Herczeg}, \& {Quijano-Vodniza}}]{2017ApJ...835....8T}
{Tofflemire}, B.~M., {Mathieu}, R.~D., {Ardila}, D.~R., {et~al.} 2017{\natexlab{a}}, \apj, 835, 8, \dodoi{10.3847/1538-4357/835/1/8}

\bibitem[{{Tofflemire} {et~al.}(2017{\natexlab{b}}){Tofflemire}, {Mathieu}, {Herczeg}, {Akeson}, \& {Ciardi}}]{2017ApJ...842L..12T}
{Tofflemire}, B.~M., {Mathieu}, R.~D., {Herczeg}, G.~J., {Akeson}, R.~L., \& {Ciardi}, D.~R. 2017{\natexlab{b}}, \apjl, 842, L12, \dodoi{10.3847/2041-8213/aa75cb}

\bibitem[{{Venuti} {et~al.}(2015){Venuti}, {Bouvier}, {Irwin}, {Stauffer}, {Hillenbrand}, {Rebull}, {Cody}, {Alencar}, {Micela}, {Flaccomio}, \& {Peres}}]{2015A&A...581A..66V}
{Venuti}, L., {Bouvier}, J., {Irwin}, J., {et~al.} 2015, \aap, 581, A66, \dodoi{10.1051/0004-6361/201526164}

\bibitem[{{Weilbacher} {et~al.}(2020){Weilbacher}, {Palsa}, {Streicher}, {Bacon}, {Urrutia}, {Wisotzki}, {Conseil}, {Husemann}, {Jarno}, {Kelz}, {P{\'e}contal-Rousset}, {Richard}, {Roth}, {Selman}, \& {Vernet}}]{2020A&A...641A..28W}
{Weilbacher}, P.~M., {Palsa}, R., {Streicher}, O., {et~al.} 2020, \aap, 641, A28, \dodoi{10.1051/0004-6361/202037855}

\bibitem[{{Whelan} {et~al.}(2014){Whelan}, {Alcal{\'a}}, {Bacciotti}, {Nisini}, {Bonito}, {Antoniucci}, {Stelzer}, {Biazzo}, {D'Elia}, \& {Ray}}]{2014A&A...570A..59W}
{Whelan}, E.~T., {Alcal{\'a}}, J.~M., {Bacciotti}, F., {et~al.} 2014, \aap, 570, A59, \dodoi{10.1051/0004-6361/201424067}

\bibitem[{{White} \& {Basri}(2003)}]{2003ApJ...582.1109W}
{White}, R.~J., \& {Basri}, G. 2003, \apj, 582, 1109, \dodoi{10.1086/344673}

\bibitem[{{White} \& {Ghez}(2001)}]{2001ApJ...556..265W}
{White}, R.~J., \& {Ghez}, A.~M. 2001, \apj, 556, 265, \dodoi{10.1086/321542}

\bibitem[{{Wilson} {et~al.}(2022){Wilson}, {Matt}, {Harries}, \& {Herczeg}}]{2022MNRAS.514.2162W}
{Wilson}, T.~J.~G., {Matt}, S., {Harries}, T.~J., \& {Herczeg}, G.~J. 2022, \mnras, 514, 2162, \dodoi{10.1093/mnras/stac1397}

\bibitem[{{Woitas} {et~al.}(2001){Woitas}, {Leinert}, \& {K{\"o}hler}}]{2001A&A...376..982W}
{Woitas}, J., {Leinert}, C., \& {K{\"o}hler}, R. 2001, \aap, 376, 982, \dodoi{10.1051/0004-6361:20011034}

\bibitem[{{Xie} {et~al.}(2020){Xie}, {Haffert}, {de Boer}, {Kenworthy}, {Brinchmann}, {Girard}, {Snellen}, \& {Keller}}]{2020A&A...644A.149X}
{Xie}, C., {Haffert}, S.~Y., {de Boer}, J., {et~al.} 2020, \aap, 644, A149, \dodoi{10.1051/0004-6361/202038242}

\bibitem[{{Zhu} {et~al.}(2024){Zhu}, {Stone}, \& {Calvet}}]{2024MNRAS.528.2883Z}
{Zhu}, Z., {Stone}, J.~M., \& {Calvet}, N. 2024, \mnras, 528, 2883, \dodoi{10.1093/mnras/stad3712}

\bibitem[{{Zsidi} {et~al.}(2022){Zsidi}, {Manara}, {K{\'o}sp{\'a}l}, {Hussain}, {{\'A}brah{\'a}m}, {Alecian}, {B{\'o}di}, {P{\'a}l}, \& {Sarkis}}]{2022A&A...660A.108Z}
{Zsidi}, G., {Manara}, C.~F., {K{\'o}sp{\'a}l}, {\'A}., {et~al.} 2022, \aap, 660, A108, \dodoi{10.1051/0004-6361/202142203}

\end{thebibliography}
\bibliographystyle{aasjournal}

\appendix

\section{Observing log}\label{App:A}
We report in Table \ref{tab:obs} the log of the MUSE observations of HT Lup. DIT stands for data integration time. FWHM-X and FWHM-Y stands for the size of the autoguider FWHM, a proxy for the line-of-sight seeing. The Strehl ratio correponds to the mean Strehl of the observation as provided in each cube header. The exposure at UT 02:10:28 and 02:51:10 on July 22, 2021 has incoherent measured Strehl values. The marked datasets corresponding to UT 02:12:01 and 02:30:25 on July 22, 2021 showcased strong saturation effects. Although HT Lup B is detected in both cubes after data reduction, they were excluded of the subsequent analysis to avoid any effect due to residual features due to this effect.

\newpage
\begin{deluxetable}{cccccccccc}[th]
\label{tab:obs}
\tablecaption{Observing log} 
\tablecolumns{8}
\tablehead{
\colhead{Date} & \colhead{UT Start} &  \colhead{DIT (s)}  & \colhead{Airmass}  &  \colhead{Strehl ratio ($\%$)} &  \colhead{FWHM-X (")} &  \colhead{FWHM-Y (")} &  \colhead{$\theta\:(^{\circ})$}}
\startdata
2021-03-25 & 08:49:56 & 20 & 1.02  & 49.82 & 0.63 & 0.64 & 0  \\ 
-- & 08:52:49 & 20 & 1.03 & 52.19 & 0.65 & 0.67 & 90 \\
-- & 08:55:55 & 90 & 1.03 & 52.17 & 0.65 & 0.67 & 0 \\
-- & 08:59:30 & 90 & 1.03 & 42.69 & 0.69 & 0.72 & 45\\
-- & 09:03:05 & 90 & 1.03 & 37.89 & 0.70 & 0.75 & 90 \\
-- & 09:06:42 & 90 & 1.04 & 33.6 & 0.68 & 0.69 & 135 \\
-- & 09:10:18 & 90 & 1.04 & 43.7 & 0.67 & 0.68 & 180 \\
-- & 09:13:53 & 90 & 1.04 & 33.57 & 0.75 & 0.79 & 225 \\
-- & 09:17:30 & 90 & 1.05 & 39.99 & 0.74 & 0.77 & 270 \\
-- & 09:24:00 & 20 & 1.05 & 30.97 & 0.74 & 0.77 & 315 \\
-- & 09:25:34 & 150 & 1.06 & 23.18 & 0.78 & 0.84 & 0 \\
2021-07-22 & 01:34:57 & 20 & 1.05  & 0.3 & 0.75 & 0.76 & 0 \\
-- & 01:37:51 & 20 & 1.05  & 17.58 & 0.70 & 0.74 & 90 \\
-- & 01:41:07 & 90 & 1.06  & 14.45 & 0.70 & 0.69 & 0  \\
-- & 01:44:42 & 90 & 1.06  & 23.96 & 0.72 & 0.73 & 45 \\
-- & 01:48:18 & 90 & 1.06  & 8.51 & 0.71 & 0.70 & 90 \\
-- & 01:51:54 & 90 & 1.07  & 13.07 & 0.70 & 0.72 & 135 \\
-- & 01:55:30 & 90 & 1.07  & 27.2 & 0.72 & 0.74 & 180 \\
-- & 01:59:05 & 90 & 1.08  & 32.76 & 0.73 & 0.70  & 225 \\
-- & 02:03:00 & 90 & 1.08  & 27.3 & 0.72 & 0.69 & 270 \\
-- & 02:06:35 & 90 & 1.09  & 31.86 & 0.70 & 0.69 & 315 \\
-- & 02:10:28 & 20 & 1.10  & \dots & 0.70 & 0.71 & 0 \\
-- & 02:12:01* & 150 & 1.10  & 30.83 & 0.71 & 0.71 & 0 \\
2021-07-22 & 02:24:11 & 20 & 1.12  & 20.95 & 0.73 & 0.69 & 0\\
-- & 02:27:04 & 20 & 1.13  & 20.73& 0.74 & 0.76 & 90 \\
-- & 02:30:25* & 90 & 1.13 & 15.36 & 0.73 & 0.75 & 0  \\
-- & 02:34:00 & 90 & 1.14 & 22.94 & 0.75 & 0.77 & 45  \\
-- & 02:37:37 & 90 & 1.15 & 18.88 & 0.75 & 0.76 & 90  \\
-- & 02:41:13 & 90 & 1.16 & 25.59 & 0.73 & 0.74 & 135  \\
-- & 02:51:10 & 90 & 1.18 & \dots & 0.91 & 0.72 & 180  \\
-- & 03:07:49 & 90 & 1.23 & 18.94 & 0.75 & 0.75 & 180  \\
-- & 03:11:24 & 90 & 1.25 & 10.94 & 0.76 & 0.78 & 225  \\
-- & 03:14:59 & 90 & 1.26 & 15.67 & 0.77 & 0.74 & 270  \\
-- & 03:18:34 & 90 & 1.27 & 16.46 & 0.73 & 0.72 & 315  \\
-- & 03:22:29 & 20 & 1.28 & 15.93 & 0.76 & 0.75 & 0  \\
-- & 03:24:02 & 150 & 1.29 & 20.11 & 0.77 & 0.73 & 0  \\
\enddata
\end{deluxetable}
\newpage
\section{HTLup B Orbital Parameters}\label{App:B}
Based on the relative astrometry of HT Lup B with respect to HT Lup A, on different epochs as reported in Table \ref{tab:sep}, we characterize its orbital parameters by using the OFTI module from the orbitize! python package \citep{2017AJ....153..229B}. This module generates random orbital parameters from prior distributions, and rejects or accepts orbits as valid by calculating the probability of the generated orbit astrometry at the epochs of observation, given the real measured astrometry and uncertainties. This process is repeated until reaching a desired number of accepted orbits.

\begin{figure*}[h!]
\epsscale{1}
\plotone{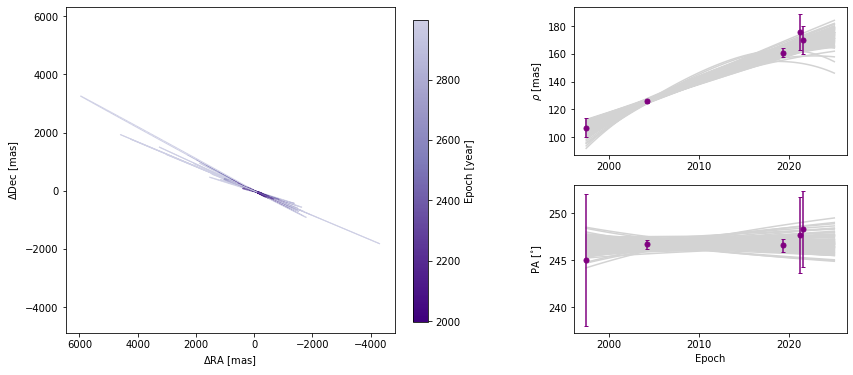}
\caption{Left: Possible orbits for HT Lup B around HT Lup A, generated with the orbitize! python package considering the five astrometry points presented in Table \ref{tab:sep}. Right: separation ($\rho$) and position angles (PA) for the retrieved orbits at different epochs, compatible with the available astrometry for HT Lup B}
\label{fig:orbits}
\end{figure*}

\begin{figure*}[h]
\epsscale{1}
\plotone{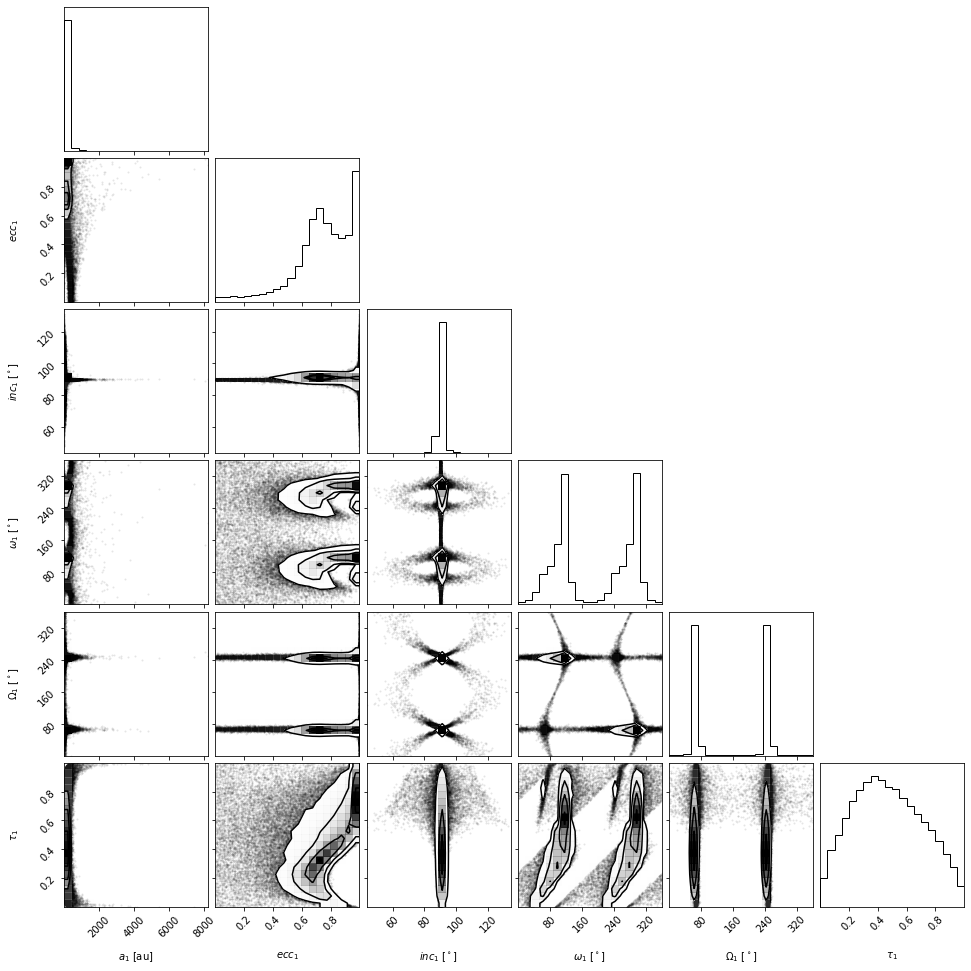}
\caption{Corner plot for the HT Lup B orbital paramaters from the orbitize! python package, considering the five astrometry points presented in Table \ref{tab:sep}. }
\label{fig:corner}
\end{figure*}

For HT Lup B, we generate a total of $10^5$ orbits. Figure \ref{fig:orbits} shows the accepted orbits, while Figure \ref{fig:corner} corresponds to the correlation between the generated parameters. Given the small orbital coverage from the available literature, most of the generated parameters do not converge to a given value, with some of then even presenting a bimodal distribution. However, it is still possible to derive some information regarding the HTLup B orbit. In particular, the inclination value of the orbit appears to strongly converge to $\approx 90^{\circ}$, indicating an almost perpendicular orbit to the on-sky plane. 

\newpage

\newpage
\section{HT Lup B Line properties} \label{App:C}
We report in Table \ref{tab:lines} the parameters derived from the characterization of the retrieved H$\alpha$ emission for HT Lup B. The analysis to derive these values is discussed in Sect. \ref{subsec:lineprop}, and was applied for each available cube individually, to check for variations on the line profile in each dataset as well as between epochs. 

\begin{deluxetable}{ccccc}[h]
\label{tab:lines}
\tablecaption{HT Lup B H$\alpha$ line properties} 
\tablecolumns{5}
\tablehead{
\colhead{Date} & \colhead{UT Start} &  \colhead{10\% linewidth} &  \colhead{50\% linewidth} & \colhead{Apparent  flux}\\
 &  &  \colhead{($km \: s^{-1}$)} &  \colhead{($km \: s^{-1}$)} & \colhead{($10^{-16} erg.s^{-1}.\ cm^{-2}$)}}
 
\startdata
2021-03-22 & 08:49:56 & $467.3 \pm 3.6$ & $247.7 \pm 9.3$ & $2.0 \pm 0.02 \times10^{3}$    \\
-- & 08:52:49 & $472.8 \pm 10.7$ & $254.3 \pm 8.1$ & $2.0 \pm 0.02 \times10^{3}$    \\
-- & 08:55:55 & $519.3 \pm 4.3$ & $283.8 \pm 6.6$ & $2.4 \pm 0.02 \times10^{3}$    \\
-- & 08:59:30 & $490.6 \pm 5.2$ & $250.8\pm 12.8$ & $2.4 \pm 0.03 \times10^{3}$  \\
-- & 09:03:05 & $487.9 \pm 4.9$ & $254.3 \pm 9.5$ & $2.2 \pm 0.02 \times10^{3}$ \\
-- & 09:06:41 & $473.5 \pm 9.3$ & $240.0 \pm 26.5$ & $2.3 \pm 0.06 \times10^{3}$  \\
-- & 09:10:17 & $486.7 \pm 4.7$ & $255.5\pm 12.1$ & $2.1 \pm 0.03 \times10^{3}$ \\
-- & 09:13:05 & $470.1 \pm 4.2$ & $243.4 \pm 10.3$ & $1.8 \pm 0.02 \times10^{3}$ \\
-- & 09:17:30 & $476.9 \pm 5.6$ & $253.3 \pm 11.4$ & $1.9 \pm 0.02 \times10^{3}$ \\
-- & 09:21:07 & $453.0 \pm 15$ & $243.3 \pm 33.3$ & $1.7 \pm 0.08 \times10^{3}$ \\
-- & 09:24:00 & $476.9 \pm 12.2$ & $226.9 \pm 34.7$ & $1.8 \pm 0.06 \times10^{3}$ \\
-- & 09:25:34 & $475.9 \pm 13.6$ & $246.4 \pm 31.4$ & $2.1 \pm 0.07 \times10^{3}$ \\
2021-07-22 & 01:34:57 & $361.2 \pm 16.4$ & $194.7\pm 20$ & $6.0 \pm 0.3 \times10^{2}$\\
-- & 01:37:51 & $413.4 \pm 10.9$ & $200.6 \pm 10.6$ & $6.7 \pm 0.2 \times10^{2}$\\
-- & 01:41:07 & $419.7 \pm 12.3$ & $190.3 \pm 5.6$ & $7.0 \pm 0.1 \times10^{2}$ \\
-- & 01:44:42 & $423.1 \pm 33.6$ & $208.3 \pm 13.8$ & $6.0 \pm 0.3 \times10^{2}$ \\
-- & 01:48:18 & $449.4 \pm 18.8$ & $196.5 \pm 11.4$ & $6.6 \pm 0.2 \times10^{2}$\\
-- & 01:51:54 & $403.6 \pm 12.9$ & $186.4 \pm 8.9$ & $6.6 \pm 0.2 \times10^{2}$\\
-- & 01:55:30 & $423.5 \pm 26.9$ & $189.6 \pm 23$ & $6.9 \pm 0.4 \times10^{2}$\\
-- & 01:59:05 & $408.4 \pm 10.1$ & $193.5 \pm 6.6$ &  $7.5 \pm 0.1 \times10^{2}$\\
-- & 02:03:00 & $420.8 \pm 17.8$ & $198.3 \pm 19.5$ &  $7.2\pm 0.4 \times10^{2}$\\
-- & 02:06:35 & $452.9 \pm 32.9$ & $194.7 \pm 22.8$  & $7.6 \pm 0.4 \times10^{2}$\\
-- & 02:10:28 & $451.7 \pm 20.9$ & $180.0 \pm 10.9$ & $6.6 \pm 0.2 \times10^{2}$\\
2021-07-22 & 02:24:11 & $454.4 \pm 55.4$ & $191.0 \pm 26.8$ & $6.4 \pm 0.4 \times10^{2}$\\
-- & 02:27:04 & $389.7 \pm 21.6$ & $193.5 \pm 11$ & $6.1 \pm 0.2 \times10^{2}$\\
-- & 02:34:00 & $400.5 \pm 22.3$ & $189.98 \pm 8.9$ & $5.7 \pm 0.2 \times10^{2}$ \\
-- & 02:37:37 & $432.5 \pm 12.7$ & $190.9 \pm 12.8$ & $6.2 \pm 0.2 \times10^{2}$\\
-- & 02:41:13 & $399.0 \pm 14.7$ & $191.8 \pm 8.5$ & $6.5 \pm 0.2 \times10^{2}$\\
-- & 03:07:49 & $400.8 \pm 21.5$ & $182.0 \pm 19.9$ & $6.0 \pm 0.3 \times10^{2}$\\
-- & 03:11:24 & $430.1 \pm 22.8$ & $199.6 \pm 21.7$ & $6.6\pm 0.3 \times 10^{2}$\\
-- & 03:14:59 & $396.8 \pm 13.0$ & $193.8 \pm 8.8$ & $5.7 \pm 0.2 \times 10^{2}$\\
-- & 03:18:35 & $416.5 \pm 27.3$ & $191.4 \pm 12.4$ & $6.1 \pm 0.3 \times 10^{2}$\\
-- & 03:22:29 & $450.4 \pm 45.9$ & $200.4 \pm 13$ & $7.3 \pm 0.3 \times 10^{2}$\\
-- & 03:24:02 & $424.2 \pm 21.2$ & $186.4 \pm 23.7$ & $6.8\pm 0.3 \times 10^{2}$\\
\enddata
\tablecomments{H$\alpha$ line properties for all available observations. All values were derived following the prescription described in Sect. \ref{subsec:lineprop}}. The reported flux values correspond to the ones directly extracted from the data after flux loss correction due to the low Strehl and over-subtraction, but without correcting for extinction.
\end{deluxetable} 

\newpage

\section{Flux loss calibration }\label{App:D}
Given the proximity of HT Lup A and B and the low Strehl ratio of MUSE for all observations, is possible that part of the light of HT Lup B will leak into the primary star, thus impacting the building of the reference spectra used for the subtraction of the stellar halo and resulting in varying levels of over and self subtraction. To properly measure this flux loss we followed a standard fake source injection approach, which allows to compare the flux of an artificial source injected into the data cubes before and after the halo removal step. The injected source was created based on the normalized stellar PSF, although the normalization was performed within a fixed aperture size, equivalent to the one used to retrieve the HT Lup B line. This allows us to ignore possible effects due to the low Strehl ratio, which are treated separatedly. The injected spectrum consisted of a Gaussian line with a FWHM and a peak flux level equivalent to that of HT Lup B in the first cube of each dataset in the case of H$\alpha$. For the [OI]$\lambda$6300 line, we instead opted to inject an artifical source with an spectrum based on the MUSE LSF at that wavelength, as the HT Lup lines are only marginally resolved and thus don't constitute a reliable spectrum model. This procedure was repeated on a sample from the first and second epoch consisting on 8 cubes each, allowing to determine if the cubes require a case by case correction or if there is an overall trend that could be applied similarly to all cubes.

\begin{figure}
\gridline{\fig{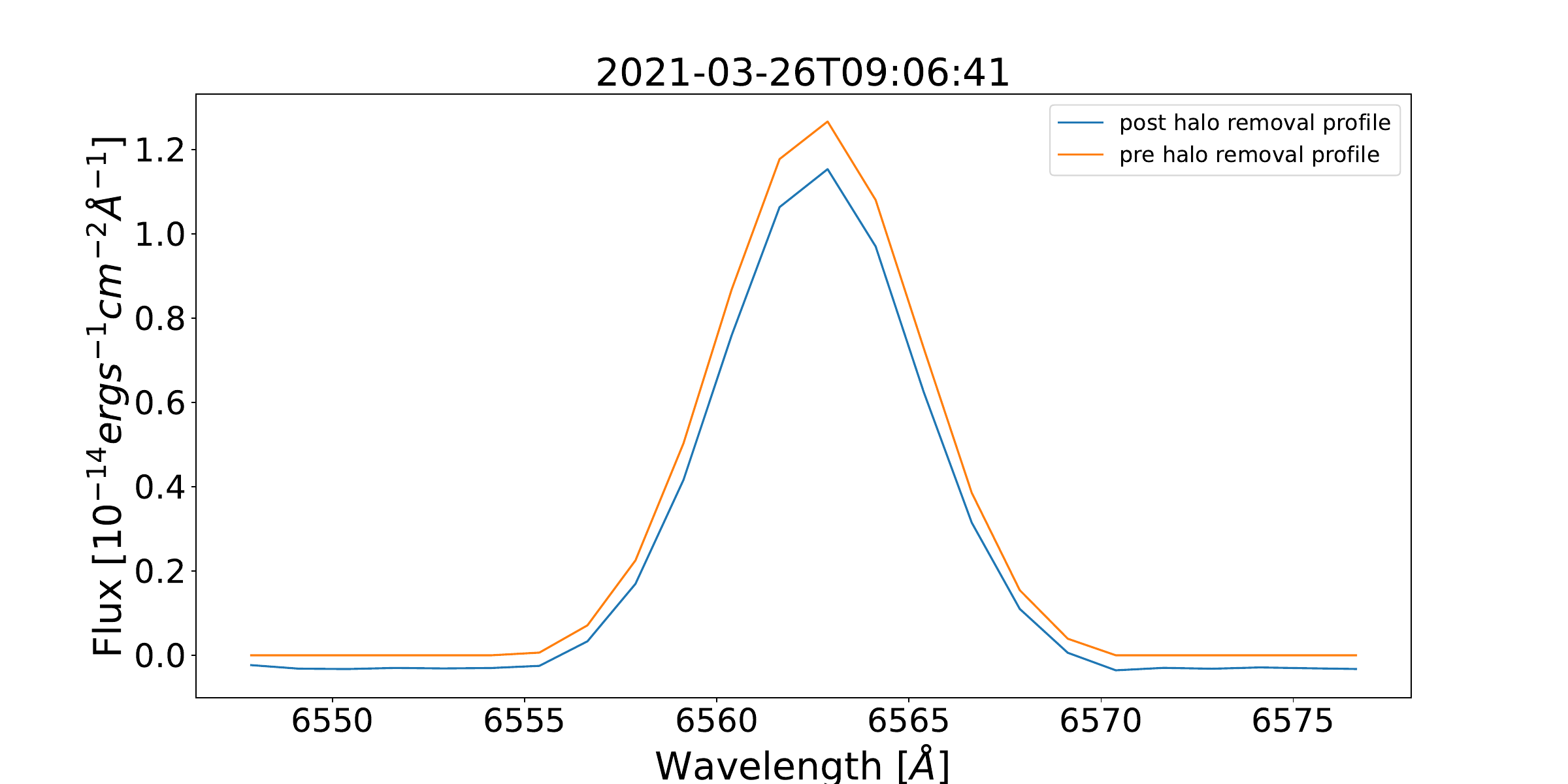}{0.45\textwidth}{(A)}
\fig{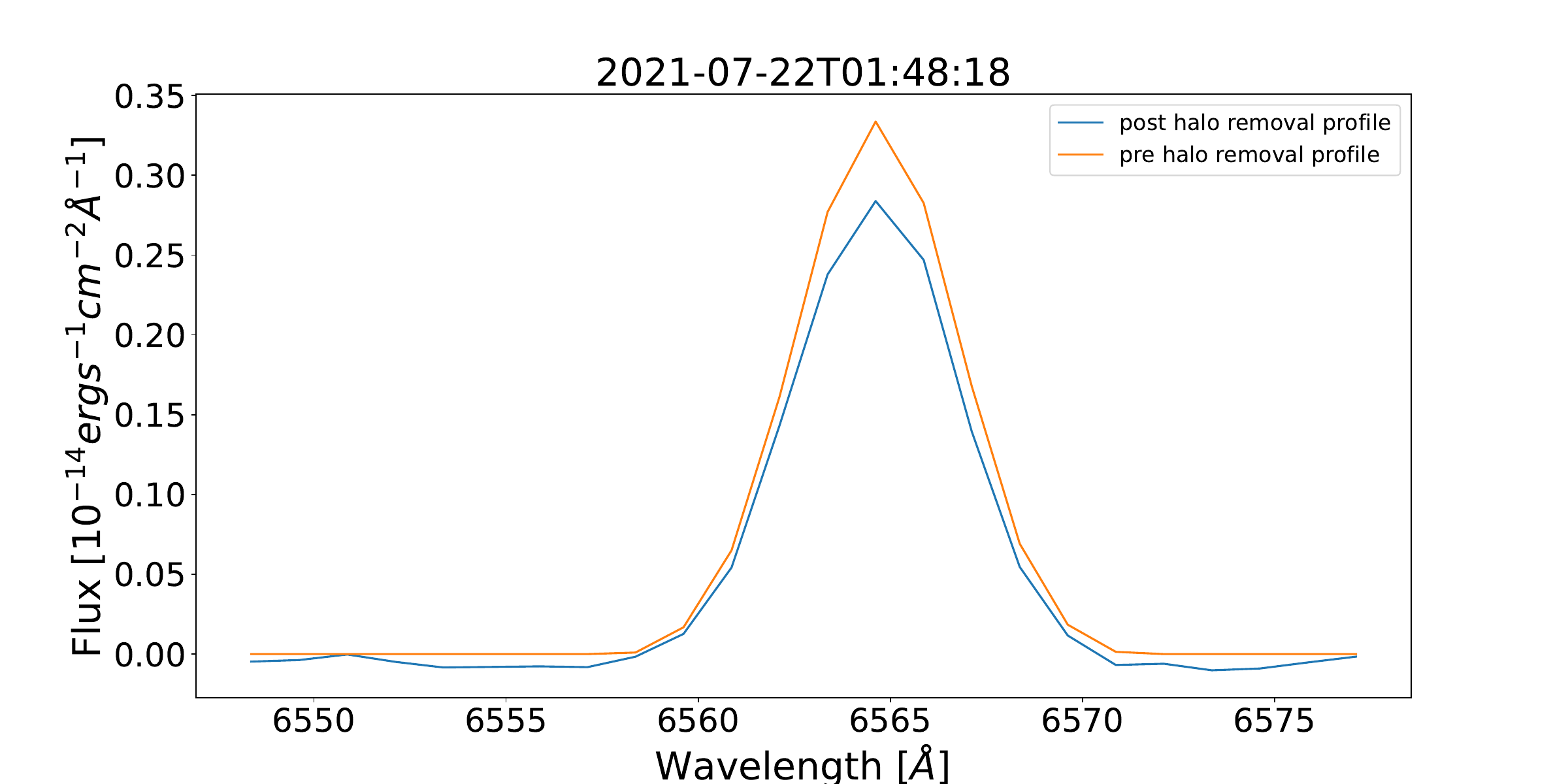}{0.45\textwidth}{(B)}}
\caption{Example of the flux variation for an injected fake source with a gaussian H$\alpha$ line profile (A) Result for one cube of the first epoch. (B) Same as (A) but for a cube of the second epoch. For each case the line was modeled so as to have a similar flux to the one of HT Lup B at the respective epoch, to better estimate the impact on a comparable source. For both cases, the post-processsed profiles remain almost identical to the pre-processed ones, with an almost constant over-subtraction along the full wavelength range being clearly identifiable on the line wings.}
\label{fig:fluxloss}
\end{figure}

\begin{figure}
\gridline{\fig{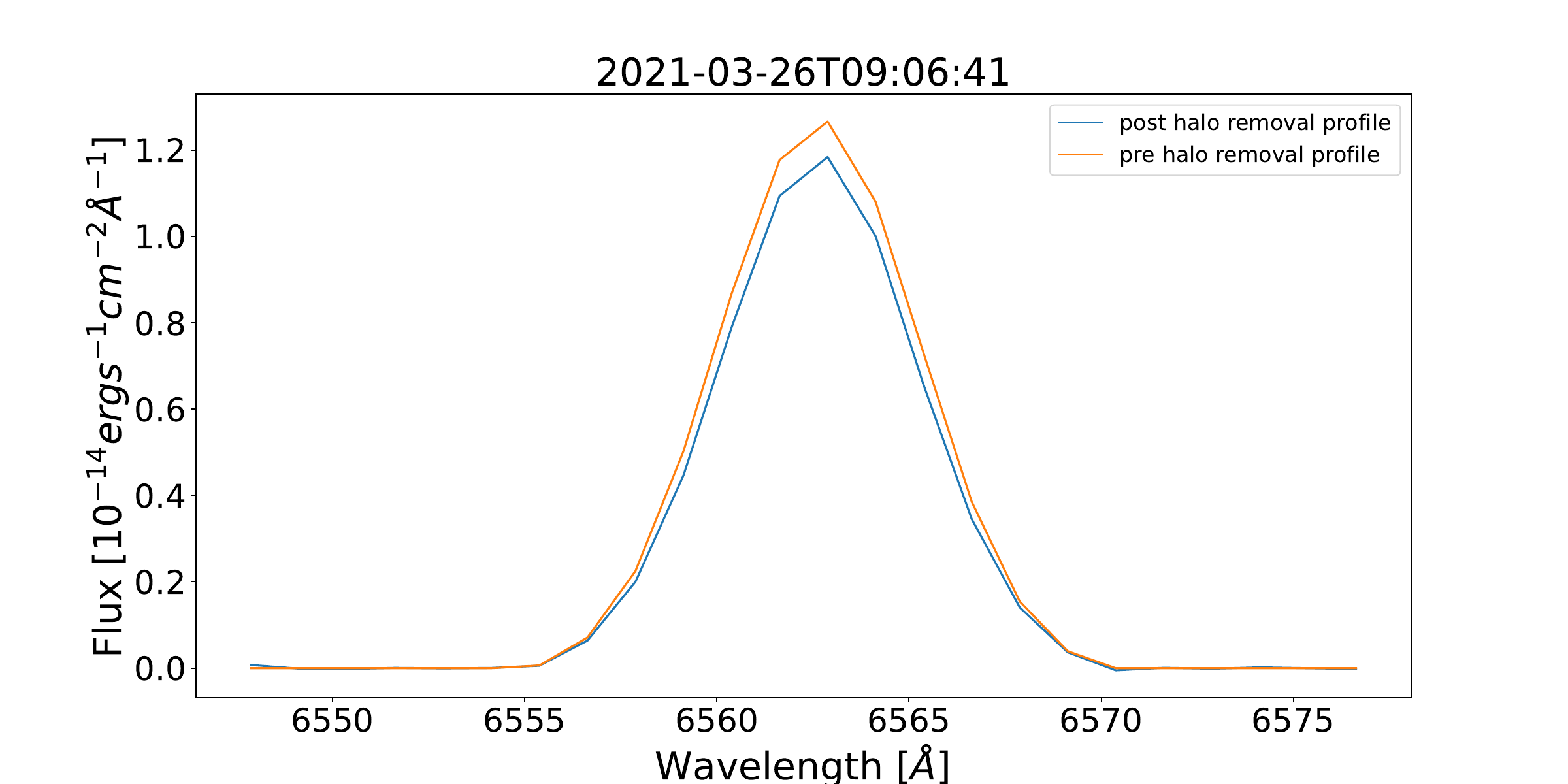}{0.45\textwidth}{(A)}
\fig{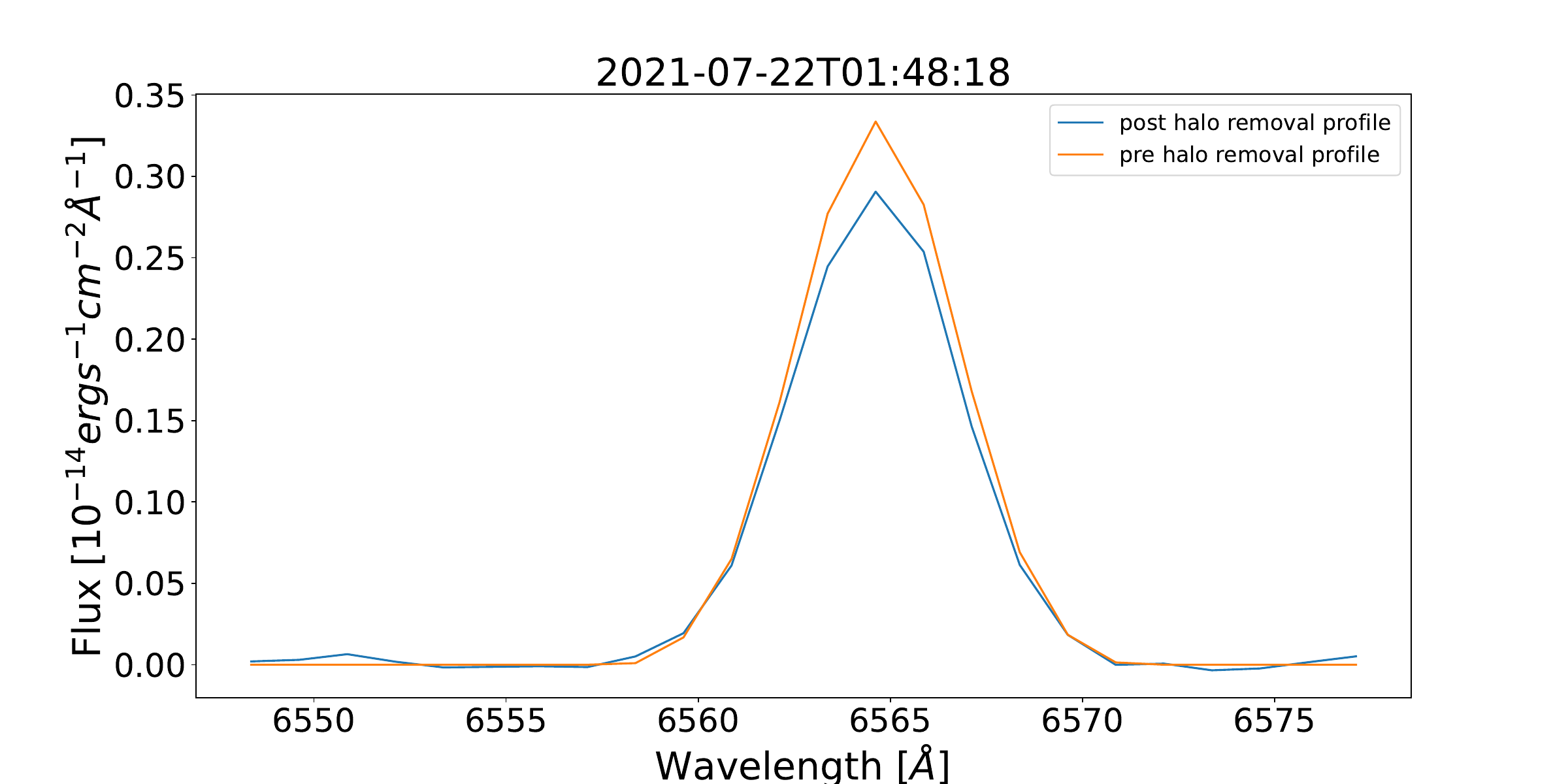}{0.45\textwidth}{(B)}}
\caption{Same as Fig.\ref{fig:fluxloss} but after corrected for the constant flux loss observed initially.}
\label{fig:fluxloss_corrected}
\end{figure}

Fig. \ref{fig:fluxloss} illustrates a representative case of the impact of the reduction on the H$\alpha$ line profile for one of the cubes of the first and second epoch. We identify that the post-processed line presents a flux loss constant with wavelength, easily identifiable by comparing the flux levels of the wings of the lines. We proceeded to estimate this flux loss by fitting a Gaussian profile plus a constant base flux level to the line. The absolute value of this base flux level was then summed to the flux at each wavelength, resulting on a final line profile as the one showcased on Fig. \ref{fig:fluxloss_corrected}.  This procedure was repeated for each sample and all the cubes showcased the same effect and similar results. After this correction, the flux difference between the pre and post-processed line flux is only a few percent, thus we do not apply any further corrections at this step. We adopted this flux correction for all the HT Lup B  H$\alpha$ lines at both epochs.

\begin{figure}[h]
\gridline{\fig{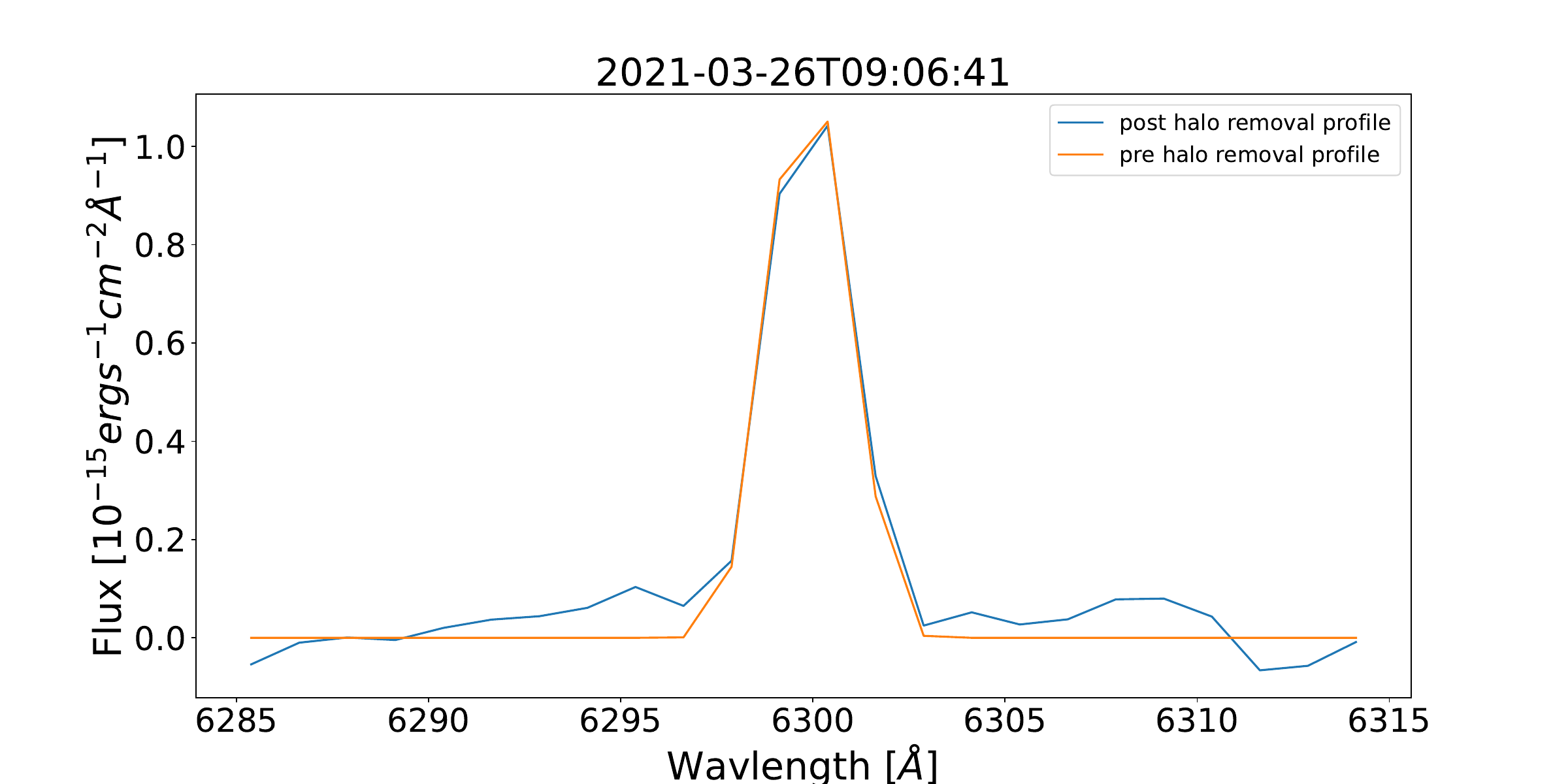}{0.45\textwidth}{(A)}
\fig{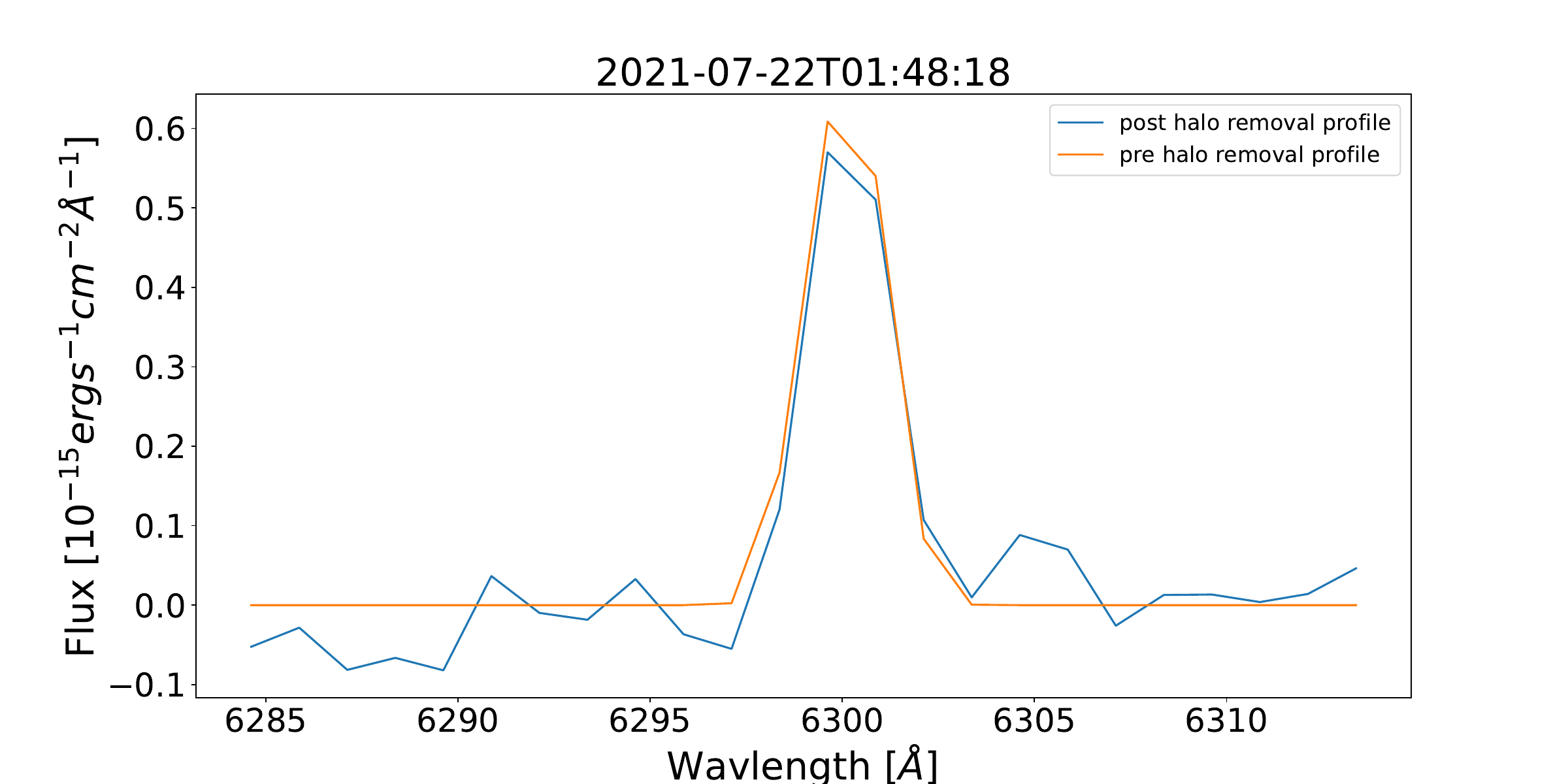}{0.45\textwidth}{(B)}}
\caption{Example of the flux variation for an injected fake source with an [OI] line profile based on the instrumental LSF. (A) Result for one cube of the first epoch. (B) Same as (A) but for a cube of the second epoch. For each case the line was modeled so as to have a similar flux to the one of HT Lup B at the respective epoch, to better estimate the impact on a comparable source. For both cases, the post-processsed profiles remain almost identical to the pre-processed ones.}
\label{fig:fluxlossOI}
\end{figure}

Fig. \ref{fig:fluxlossOI} showcases a representative case of the impact of the reduction on the [OI]$\lambda$6300 line profile for one of the cubes of the first and second epoch. Opposite to the case of H$\alpha$, the flux loss on this case is already only a few percent without any correction. As we only report the cases with SNR>10 which correlate to the cubes less affected by self-subtractions based on our fake source injections, we opted to not apply any further calibration to the lines at this step.

Given the low Strehl ratio of the observations, the use of an small aperture with a size similar to the FWHM of the PSF could suffer from significant flux loss. As the Strehl ratio can also be affected due to variations on the observing conditions, is possible that part of the observed variability of the lines is a product of this effect. To quantify and correct for this flux loss, we calculated the flux ratio of HT Lup A between an aperture with a radius of 3.5 pixels as the one used for HT Lup B and an aperture with a radius of 1" for all cubes. For the particular case of H$\alpha$ we calculated the ratio at a close wavelength region but avoiding the line to both minimize the impact of HT Lup B as well as avoid any discrepancy due to the different line profile of HT Lup A. We repeated this procedure for all the retrieved lines, namely H$\alpha$, H$\beta$, and [OI]$\lambda$6300 and for all the cubes of the sample.

\newpage
\section{HT Lup A spectrum and emission line retrieval}\label{App:E}
\begin{figure*}[th!]
\plotone{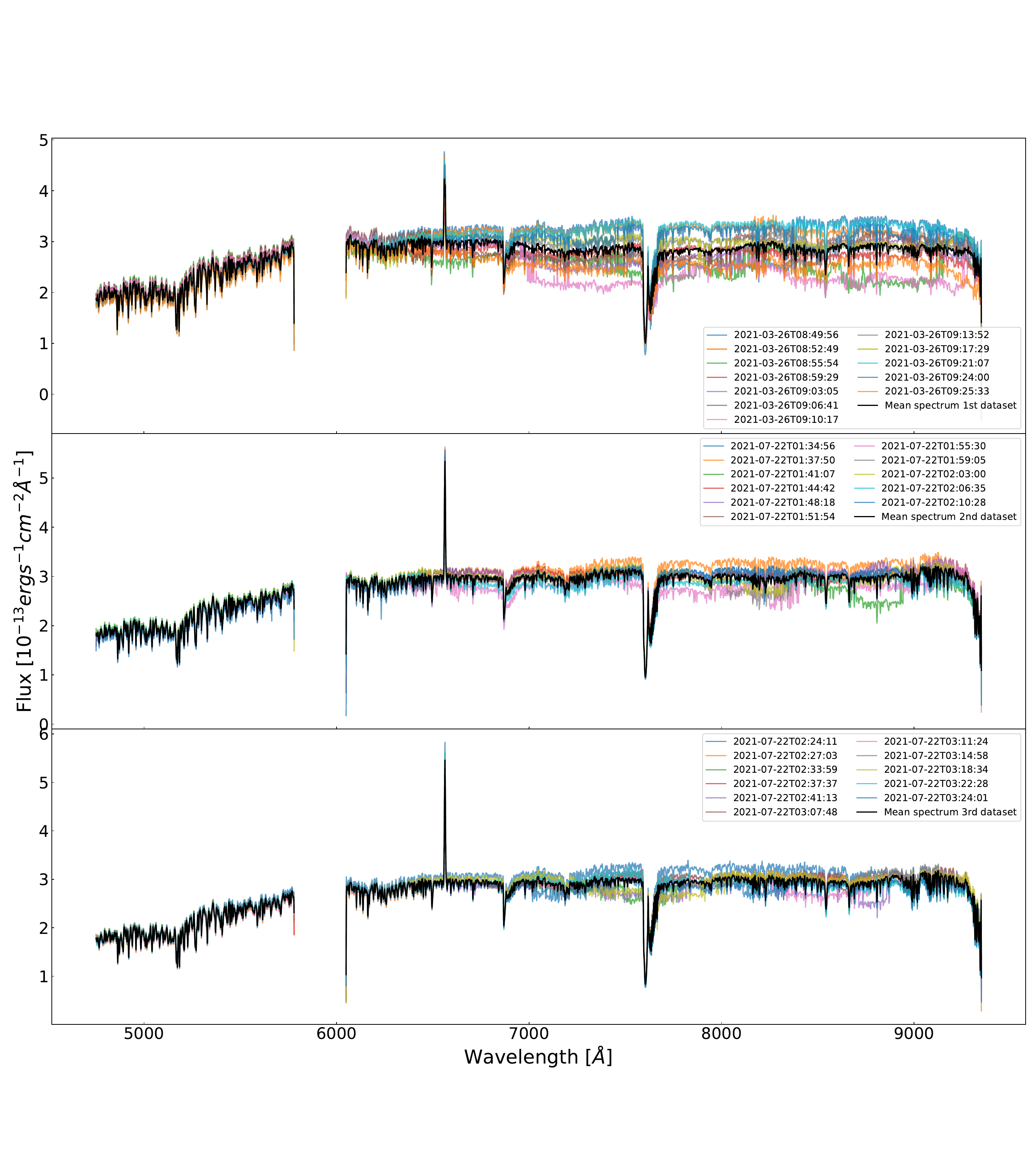}
\caption{Spectra of HT Lup A retrieved for each of the available datacubes. Top, middle and bottom panels correspond to the first, second and third dataset respectively. The black line in each panel corresponds to the mean spectrum for each dataset. Besides evident H$\alpha$ emission, HT Lup A does not appear to exhibit any other lines associated with accretion.} 
\label{fig:spectra}
\end{figure*}

For each individual observation of the three available datasets, the spectrum of HT Lup A was extracted within a circular aperture on the wavelength-calibrated, recentered cubes. The center of the emission at each wavelength, and thus of the aperture, was determined by fitting a 2D elliptic Moffat profile to the wings of the PSF. To avoid the effect of the low Strehl ratio of MUSE on small apertures, the flux was retrieved using an aperture radius of 40 pixels (1"). The final spectra for each dataset is presented in Fig. \ref{fig:spectra} 

The next step was to characterize the emission lines associated with accretion. For all datasets, the HT Lup A spectra lacks any accretion signature besides H$\alpha$ line emission, thus the analysis only focuses on that line. To isolate the H$\alpha$ emission line from one observation, we first extract a sub-region of the spectrum between 6400-6700 \AA to model the continuum emission of the source near H$\alpha$. This is done by first smoothing the spectra using a median filter with a window size of 27 elements, which is then interpolated using a 3rd order polynomial. The modeled continuum is then subtracted from the sub-spectrum, retaining only the flux corresponding to the emission line. Finally, we substract the contribution from HT Lup B at each wavelength by removing the corresponding emission retrieved after the removal of the stellar halo and flux loss correction. This procedure is repeated for each cube of each dataset. We follow the same procedure presented in Sect. \ref{subsec:lineprop} to retrieve the line parameters, namely the line width, peak velocity and total apparent flux, which are presented in Table \ref{tab:lineshtlupA}

\begin{deluxetable}{ccccc}[h]
\label{tab:lineshtlupA}
\tablecaption{HT Lup A H$_{\alpha}$ line properties} 
\tablecolumns{5}
\tablehead{
\colhead{Date} & \colhead{UT Start} &  \colhead{10\% linewidth} &  \colhead{50\% linewidth} & \colhead{Apparent  flux}\\
 &  &  \colhead{($km \: s^{-1}$)} &  \colhead{(km/s)} & \colhead{($10^{-16} erg.s^{-1}.\ cm^{-2}$)}}
 
\startdata
2021-03-22 & 08:49:56 & $535 \pm 124$ & $403.4 \pm 17.1$ & $1.09 \pm 0.04 \times10^{4}$   \\
-- & 08:52:49 & $553.2 \pm 95.9$ & $412.4 \pm 18.7$ &  $1.20 \pm 0.05 \times10^{4}$   \\
-- & 08:55:55 & $567.2 \pm 86.6$ & $400.4 \pm 20.6$ &  $9.99 \pm 0.37 \times10^{3}$   \\
-- & 08:59:30 & $523 \pm 114$ & $412.2 \pm 10.2$ & $1.25 \pm 0.05 \times10^{4}$  \\
-- & 09:03:05 & $540 \pm 119$ & $385.5 \pm 29.8$ & $1.17 \pm 0.04 \times10^{4}$  \\
-- & 09:06:41 & $616.6 \pm 59.6$ & $417.0 \pm 23.9$ & $1.11 \pm 0.05 \times10^{4}$  \\
-- & 09:13:05 & $541.3 \pm 77.3$ & $406.8 \pm 12.4$ & $1.32 \pm 0.04 \times10^{4}$ \\
-- & 09:17:30 & $567 \pm 130$ & $397.2 \pm 33$ & $1.34 \pm 0.06 \times10^{4}$ \\
-- & 09:21:07 & $527 \pm 132$ & $392 \pm 26$ & $1.22 \pm 0.05 \times10^{4}$ \\
-- & 09:24:00 & $543.9 \pm 96.7$ & $400.2 \pm 15.2$ & $1.23 \pm 0.05 \times10^{4}$ \\
-- & 09:25:34 & $479.36 \pm 6.52$ & $266.6 \pm 14.2$ & $9.98 \pm 0.43 \times10^{3}$ \\
2021-07-22 & 01:34:57 & $474.5 \pm 31.6$ & $256.6\pm 10.4$ & $1.59 \pm 0.05 \times10^{4}$\\
-- & 01:37:51 & $465.7 \pm 26.1$ & $263.9 \pm 10.3$ & $1.6 \pm 0.05\times10^{4}$\\
-- & 01:41:07 & $501 \pm 82$ & $274.3\pm 14.7$  & $1.37 \pm 0.05 \times10^{4}$ \\
-- & 01:44:42 & $472.5 \pm 30.7$ & $262.7 \pm 11.6$ & $1.37 \pm 0.04 \times10^{4}$ \\
-- & 01:48:18 & $461.4 \pm 21.4$ & $259.39 \pm 9.07$ & $1.49 \pm 0.04 \times10^{4}$\\
-- & 01:51:54 & $460.9 \pm 71.7$ & $258.7 \pm 15.6$ & $1.44 \pm 0.07 \times10^{4}$\\
-- & 01:55:30 & $488.2 \pm 51.5$ & $289.7 \pm 12.2$ & $1.47 \pm 0.05 \times10^{4}$\\
-- & 01:59:05 & $458.8 \pm 70.5$ & $264.9 \pm 14.2$ &  $1.37 \pm 0.06 \times10^{4}$\\
-- & 02:03:00 & $435 \pm 117$ & $257.8 \pm 26.4$ &  $1.34 \pm 0.09 \times10^{4}$\\
-- & 02:06:35 & $470 \pm 103$ & $253.5 \pm 22.5$  & $1.47 \pm 0.08 \times10^{4}$\\
-- & 02:10:28 & $464.3 \pm 27.5$ & $263.1 \pm 13.7$ & $1.45 \pm 0.05 \times10^{4}$\\
-- & 02:12:01 & $469.9 \pm 47.3$ & $258.8 \pm 15.6$  & $9.44 \pm 0.39 \times10^{3}$\\
2021-07-22 & 02:24:11 & $470.2 \pm 27.4$ & $258.4 \pm 10$ & $1.57 \pm 0.05 \times10^{4}$\\
-- & 02:27:04 & $465.1 \pm 33.9$ & $260.94 \pm 9.65$ & $1.53 \pm 0.05 \times10^{4}$\\
-- & 02:30:25 & $501.9 \pm 56.8$ & $284.6 \pm 11.1$ &  $1.57 \pm 0.04 \times10^{4}$\\
-- & 02:34:00 & $466.3 \pm 40$ & $272.3 \pm 11.8$ & $1.64 \pm 0.06 \times10^{4}$ \\
-- & 02:37:37 & $455.3 \pm 28.7$ & $261.8 \pm 10.2$ & $1.59 \pm 0.05 \times10^{4}$\\
-- & 02:41:13 & $470.5 \pm 47.3$ & $254.3 \pm 12.6$ & $1.59 \pm 0.06 \times10^{4}$\\
-- & 03:07:49 & $459.6 \pm 28$ & $260.38 \pm 9.71$ & $1.57 \pm 0.05 \times10^{4}$\\
-- & 03:11:24 & $464.3 \pm 22.9$ & $259.61 \pm 9.6$ & $1.49 \pm 0.04 \times 10^{4}$\\
-- & 03:14:59 & $425.6 \pm 25.6$ & $236.86 \pm 9.05$ & $1.41 \pm 0.05 \times 10^{4}$\\
-- & 03:18:35 & $478 \pm 21.7$ & $276.3 \pm 10$ & $1.51 \pm 0.04 \times 10^{4}$\\
-- & 03:22:29 & $457.3 \pm 30.2$ & $252.5 \pm 10.3$ & $1.62 \pm 0.05 \times 10^{4}$\\
-- & 03:24:02 & $478 \pm 40.2$ & $276.2 \pm 11.2$ & $1.44 \pm 0.05 \times 10^{4}$\\
\enddata
\tablecomments{Same as Table \ref{tab:lines}, but with H$\alpha$ line parameters derived for HT Lup A}.
\end{deluxetable}
\end{document}